\documentclass[11pt]{article}
\pdfoutput=1
\usepackage{amsfonts,amsmath,dsfont}
\usepackage{amssymb}
\usepackage{fancyhdr}
\usepackage{slashed}
\usepackage{graphicx}
\usepackage{subfigure}
\usepackage{color}
\usepackage{setspace}
\usepackage{float}
\usepackage{comment}

\usepackage{hyperref}
\usepackage[utf8]{inputenc}
\usepackage[titletoc]{appendix}

\usepackage{cleveref}

\def\hybrid{\topmargin -20pt    \oddsidemargin 0pt
        \headheight 0pt \headsep 0pt
        \textwidth 6.25in       % A4 paper
        \textheight 9.25in       % A4 paper
        \marginparwidth .875in
        \parskip 5pt plus 1pt   \jot = 1.5ex}

\hybrid

\def\baselinestretch{1.2}

\catcode`\@=11

\def\marginnote#1{}

\newcount\hour
\newcount\minute
\newtoks\amorpm
\hour=\time\divide\hour by60
\minute=\time{\multiply\hour by60 \global\advance\minute by-\hour}
\edef\standardtime{{\ifnum\hour<12 \global\amorpm={am}%
        \else\global\amorpm={pm}\advance\hour by-12 \fi
        \ifnum\hour=0 \hour=12 \fi
        \number\hour:\ifnum\minute<10 0\fi\number\minute\the\amorpm}}
\edef\militarytime{\number\hour:\ifnum\minute<10 0\fi\number\minute}
\def\draftlabel#1{{\@bsphack\if@filesw {\let\thepage\relax
   \xdef\@gtempa{\write\@auxout{\string
      \newlabel{#1}{{\@currentlabel}{\thepage}}}}}\@gtempa
   \if@nobreak \ifvmode\nobreak\fi\fi\fi\@esphack}
        \gdef\@eqnlabel{#1}}
\def\@eqnlabel{}
\def\@vacuum{}
\def\draftmarginnote#1{\marginpar{\raggedright\scriptsize\tt#1}}

\def\draft{\oddsidemargin -.5truein
        \def\@oddfoot{\sl preliminary draft \hfil
        \rm\thepage\hfil\sl\today\quad\militarytime}
        \let\@evenfoot\@oddfoot \overfullrule 3pt
        \let\label=\draftlabel
        \let\marginnote=\draftmarginnote
   \def\@eqnnum{(\theequation)\rlap{\kern\marginparsep\tt\@eqnlabel}%
\global\let\@eqnlabel\@vacuum}  }

\def\preprint{\twocolumn\sloppy\flushbottom\parindent 2em
        \leftmargini 2em\leftmarginv .5em\leftmarginvi .5em
        \oddsidemargin -.5in    \evensidemargin -.5in
        \columnsep .4in \footheight 0pt
        \textwidth 10.in        \topmargin  -.4in
        \headheight 12pt \topskip .4in
        \textheight 6.9in \footskip 0pt
        \def\@oddhead{\thepage\hfil\addtocounter{page}{1}\thepage}
        \let\@evenhead\@oddhead \def\@oddfoot{} \def\@evenfoot{} }

\def\numberbysection{\@addtoreset{equation}{section}
        \def\theequation{\thesection.\arabic{equation}}}

\def\underline#1{\relax\ifmmode\@@underline#1\else
        $\@@underline{\hbox{#1}}$\relax\fi}

\def\titlepage{\@restonecolfalse\if@twocolumn\@restonecoltrue\onecolumn
     \else \newpage \fi \thispagestyle{empty}\c@page\z@
        \def\thefootnote{\fnsymbol{footnote}} }

\def\endtitlepage{\if@restonecol\twocolumn \else \newpage \fi
        \def\thefootnote{\arabic{footnote}}
        \setcounter{footnote}{0}}  %\c@footnote\z@ }

\catcode`@=12
\relax

\def\figcap{\section*{Figure Captions\markboth
        {FIGURECAPTIONS}{FIGURECAPTIONS}}\list
        {Figure \arabic{enumi}:\hfill}{\settowidth\labelwidth{Figure
999:}
        \leftmargin\labelwidth
        \advance\leftmargin\labelsep\usecounter{enumi}}}
 \relax
\def\tablecap{\section*{Table Captions\markboth
        {TABLECAPTIONS}{TABLECAPTIONS}}\list
        {Table \arabic{enumi}:\hfill}{\settowidth\labelwidth{Table
999:}
        \leftmargin\labelwidth
        \advance\leftmargin\labelsep\usecounter{enumi}}}
 \relax
\def\reflist{\section*{References\markboth
        {REFLIST}{REFLIST}}\list
        {[\arabic{enumi}]\hfill}{\settowidth\labelwidth{[999]}
        \leftmargin\labelwidth
        \advance\leftmargin\labelsep\usecounter{enumi}}}
 \relax
\makeatletter
\newcounter{pubctr}
\def\publist{\@ifnextchar[{\@publist}{\@@publist}}
\def\@publist[#1]{\list
        {[\arabic{pubctr}]\hfill}{\settowidth\labelwidth{[999]}
        \leftmargin\labelwidth
        \advance\leftmargin\labelsep
        \@nmbrlisttrue\def\@listctr{pubctr}
        \setcounter{pubctr}{#1}\addtocounter{pubctr}{-1}}}
\def\@@publist{\list
        {[\arabic{pubctr}]\hfill}{\settowidth\labelwidth{[999]}
        \leftmargin\labelwidth
        \advance\leftmargin\labelsep
        \@nmbrlisttrue\def\@listctr{pubctr}}}
 \relax
\makeatother
\newskip\humongous \humongous=0pt plus 1000pt minus 1000pt

\newif\ifdtup

\relax

\def\be{\begin{equation}}
\def\ee{\end{equation}}
\def\ba{\begin{eqnarray}}
\def\ea{\end{eqnarray}}

\def\a{\alpha}

\def\b{\beta}

\def\g{\gamma}

\def\e{\epsilon}

\def\th{\theta}

\def\m{\mu}

\def\n{\nu}

\def\S{\Sigma}

 \def\cH{{\cal H}}

  \def\cO{{\cal O}}
\def\cP{{\cal P}}  
\def\cS{{\cal S}} \def\cT{{\cal T}}

\newcommand{\vev}[1]{{\left< {#1} \right>}}
\newcommand{\bra}[1]{{\left< {#1} \right|}}
\newcommand{\ket}[1]{{\left| {#1} \right>}}
\newcommand{\prt}[1]{{\left( {#1} \right)}}
\newcommand{\prtt}[1]{{\left[ {#1} \right]}}

\def\no{\noindent}

\def\IR{\relax{\rm I\kern-.18em R}}

\def\pp{\partial}

\newcommand{\ff}{\dfrac}

\def\IR{\relax{\rm I\kern-.18em R}}
\def\IL{\relax{\rm I\kern-.18em L}}

\def\inv{^{\raise.15ex\hbox{${\scriptscriptstyle -}$}\kern-.05em 1}}

\def\Tr{{\rm Tr}}

\def\bea{\begin{eqnarray}}
\def\eea{\end{eqnarray}}
\newcommand{\eq}[1]{(\ref{#1})}
\def\nn{\nonumber}

\newcommand{\la}[1]{\label{#1}}

\def\a{\alpha}      
\def\b{\beta}       
\def\g{\gamma}    
    
\def\e{\epsilon}

\def\m{\mu} \def\n{\nu}

  \def\S{\Sigma}
\def\t{\tau}
\def\th{\theta}

\definecolor{markcolor2}{rgb}{1,0,0}

\definecolor{markcolor3}{rgb}{0,1,0}

\begin{document}

\renewcommand{\theequation}{\thesection.\arabic{equation}}
\csname @addtoreset\endcsname{equation}{section}

\newcommand{\beq}{\begin{equation}}
\newcommand{\eeq}[1]{\label{#1}\end{equation}}
\newcommand{\ber}{\begin{eqnarray}}
\newcommand{\eer}[1]{\label{#1}\end{eqnarray}}
\newcommand{\eqn}[1]{(\ref{#1})}
\begin{titlepage}

\begin{center}

~
\vskip .7 cm

{\Large
\bf Holographic Timelike Entanglement Entropy in Non-relativistic Theories

}

%{\Large
%\bf   ..
%}

\vskip 0.6in

 {\bf Mir Afrasiar,${}^{1,2}$
	Jaydeep Kumar Basak,${}^{1,2}$ and
	Dimitrios Giataganas${}^{1,2,3}$}
 \vskip 0.1in
 {\em
 	{\it ${}^1$
		Department of Physics, National Sun Yat-Sen University,
		Kaohsiung 80424, Taiwan\\}
	{\it ${}^2$
		Center for Theoretical and Computational Physics,
		Kaohsiung 80424, Taiwan\\}
	{\it ${}^3$
		Physics Division, National Center for
		Theoretical Sciences, Taipei 10617, Taiwan\\}
~ \\~\vskip .25in
 {\tt\href{mailto:mirhepth@gmail.com}{mirhepth@gmail.com}, \href{mailto:jkb.hep@gmail.com}{jkb.hep@gmail.com},\\
	\href{mailto:dimitrios.giataganas@mail.nsysu.edu.tw}
	{dimitrios.giataganas@mail.nsysu.edu.tw}
 }\\
 }

\vskip .1in
\end{center}

\vskip .8in

\centerline{\bf Abstract}
\noindent

Timelike entanglement entropy is a complex measure of information that is holographically realized by an appropriate combination of spacelike and timelike extremal surfaces. This measure is highly sensitive to Lorentz invariance breaking. In this work, we study the timelike entanglement entropy in non-relativistic theories, focusing on theories with hyperscaling violation and Lifshitz-like spatial anisotropy. The properties of the extremal surfaces, as well as the timelike entanglement entropy itself, depend heavily on the symmetry-breaking parameters of the theory. Consequently, we show that timelike entanglement can encode, to a large extent, the stability and naturalness of the theory. Furthermore, we find that timelike entanglement entropy identifies Fermi surfaces either through the logarithmic behavior of its real part or, alternatively, via its constant imaginary part, with this constant value depending on the theory's Lifshitz exponent. This provides a novel interpretation for the imaginary component of this pseudoentropy. Additionally, we examine temporal entanglement entropy, an extension of timelike entanglement entropy to Euclidean space, and provide a comprehensive discussion of its properties in these theories.

\no
\end{titlepage}
\vfill
\eject

\newpage
\begin{spacing}{1}
\tableofcontents
\end{spacing}

\newpage
%\end{center}

\noindent

%\vskip .4in
%\noindent
%August 2002\\
%\end{titlepage}
%\vfill
%\eject

\def\baselinestretch{1.2}
\baselineskip 19 pt
\noindent

%%%%%%%%%%%%%%%

\setcounter{equation}{0}

\section{Introduction}\label{sec_intro}

In recent years, theoretical high-energy physics and quantum information theory have found common ground, particularly through studies of entanglement entropy \cite{Ryu:2006bv}. This interdisciplinary interest is motivated by the potential to describe complex systems, investigate quantum and other phase transitions, and relate structures in quantum many-body systems to the geometric properties of gravitational theories in holographic duals, among other topics. Many aspects of these ideas have been well explored in the literature already.

Recent developments have extended the concept of entanglement entropy to certain types of pseudoentropies, a notion naturally motivated by holographic correspondence. For a subregion $A$ the entanglement entropy is defined by decomposing the Hilbert space as $\cH=\cH_A\otimes\cH_B$  for $B=A^c$  and calculating the von Neumann entropy of the reduced density matrix for a quantum state $\ket{\Psi}\in \cH$. In a holographic framework, entanglement entropy corresponds to the area of the extremal surface that reaches the boundary in the static gravity dual. Conversely, certain types of pseudoentropy introduced involve two pure states, the initial state $\ket{\Psi}$ and the final one $\ket{\Phi}$ with a transition matrix defined between them. Here, the reduced density matrix is replaced by the reduced transition matrix:
%\be
$\t_A=\Tr_B\prtt{\ket{\Psi}\bra{\Phi}/\prt{\bra{\Phi}\ket{\Psi}}}~,$
%\ee
and the pseudoentropy is  given by  $S_A=- \Tr\prtt{\t_A \log \t_A}$.
We note that the reduced transition matrix is not Hermitian in general and therefore the pseudoentropy is complex valued. This non-Hermiticity is consistent with postselection setups, where operator expectation values $\vev{\cO}=\Tr\prtt{\cO \t}$, are given by the use of the transition matrix. The expectation value yields physically meaningful real and imaginary components, extensively studied in weak value contexts \cite{Aharonov:1988xu,RevModPhys.86.307}. Notably, if $\ket{\Psi}=\ket{\Phi}$ the pseudoentropy reverts to the real-valued entanglement entropy. An extensive analysis of the pseudoentropy for free field theories in the harmonic chain, quantum spin models and marginally deformed CFTs is presented in  \cite{Mollabashi:2020yie,Mollabashi:2021xsd}.

In this paper we are mainly interested in studying a potentially discrete complex-valued measure of information, a timelike entanglement entropy (tEE) which in certain setups has been proven to match the pseudoentropy or being a special example of it, as it has been defined above. A key question is how we can explicitly define and compute this tEE. A natural definition involves an analytical continuation of the entanglement entropy to a timelike subsystem $A$. This results in a tEE with a constant imaginary component in two-dimensional quantum field theories on flat spacetime, determined by the central charge. Alternatively, one may directly Wick-rotate the field theory coordinates, beginning with a Lorentz-invariant free scalar field theory to compute tEE. In two-dimensional theories, these methods agree exactly. However, such definitions may encounter challenges in theories without Lorentz invariance, such as those with non-relativistic Lifshitz symmetry and hyperscaling violation \cite{hyper1} which are of major interest. These theories describe for example diverse quantum critical phenomena \cite{Henley_1997,Ardonne:2003wa,Fradkin_2004,Vishwanath_2004,Ghaemi_2005,PhysRevLett.93.066401}, Fermi liquids, and other non-Fermi liquids \cite{Wolf:2006zzb,Gioev:2006zz,Swingle:2010yi,spinliquid}, which are of considerable experimental and theoretical interest, with extensive literature on holographic approaches, for example \cite{Kachru:2008yh,Dong:2012se,Ogawa:2011bz,Huijse:2011ef}. Applying the same definition of the tEE to these theories, leads to higher time-derivative terms, raising quantization and stability challenges.

Remarkably, the holographic dual of the tEE is still conjectured to be the area of an extremal surface that correspond to the boundary time interval $T$ and comprises both timelike and spacelike surfaces. The conjecture has been confirmed in ($2+1$)-dimensional theories \cite{Doi:2023zaf}. In AdS$_3$, the tEE can be computed by appropriately considering the union of the spacelike and timelike geodesics and computing their areas. A related quantity, the temporal entanglement entropy \cite{Grieninger:2023knz,Doi:2023zaf,Afrasiar:2024lsi}, is defined by Wick-rotating the time coordinate to obtain a Euclidean metric. In ($2+1$)-dimensional theories, temporal and timelike entanglement entropies match with an application of a Wick-rotation, though differences arise in higher dimensions. It is important to note that the temporal entanglement entropy is not equal to the tEE.

The holographic computation of tEE has the advantage of a broadly applicable and universal definition. In our approach, we consider the union of the spacelike and timelike extremal surfaces homologous to the timelike region. To connect a timelike subregion’s boundary via extremal surfaces, both spacelike and timelike extremal surfaces are required. The real and imaginary components of the tEE correspond to spacelike and timelike parts of these extremal surfaces, respectively, making tEE in general complex. This is the definition of the tEE we follow in this paper. It explicitly agrees with the holographic computations of \cite{Doi:2023zaf} when applied to theories studied there, where previously the notion of the tEE was introduced.

The tEE offers potential significant applications and insights into quantum many-body systems and field theories. It could elucidate fundamental principles of gauge/gravity correspondence, as tEE is intrinsically tied to the emergence of a time coordinate. Moreover, tEE can serve as a novel order parameter in quantum many-body systems, potentially surpassing in certain cases the entanglement entropy in sensitivity to non-relativistic symmetries. All these developments strongly motivate further study of timelike and temporal entanglement entropies in non-relativistic theories that describe for example fixed points with Lifshitz symmetry, hyperscaling violation symmetry, and spatially anisotropic Lifshitz-like symmetries. This is the aim of this manuscript.

To rigorously compute the holographic tEE, we solve for all extremal spacelike and timelike surfaces that correspond to the chosen timelike boundary interval. This can be formally done at the level of the equations of motion of these surfaces, where all the possibilities of solutions are investigated. We compute the gradient normal vector field, whose norm provides conditions for the spacelike and timelike surfaces in the bulk and where these surfaces meet, illustrating their arrangement for each theory. After determining these surfaces, we compute their extremal areas to derive the real and imaginary components of tEE, which receive contributions from the spacelike and timelike parts exclusively. This holographic approach aligns with fundamental holographic principles.

Each of the real and imaginary components of tEE are especially noteworthy and has been found to have several interesting properties. The real part can probe confinement and consequently confinement/deconfinement phase transitions \cite{Afrasiar:2024lsi}. This is in analogy with the real part of pseudoentropy in field theories which can act as an order parameter for various quantum phases \cite{Mollabashi:2020yie,Mollabashi:2021xsd}, where in particular the difference between the real part of pseudoentropy and the entanglement entropy is non-negative or non-positive if states are in different or same quantum phases, respectively. On the other hand, the imaginary part has some interesting properties as well. The imaginary component of the tEE, is constant in  ($2+1$)-dimensional static theories, and can serve as a direct measure of the central charge of the theory \cite{Doi:2023zaf,Afrasiar:2024lsi,Chu:2023zah}. In confining theories, it suddenly drops to zero once the boundary interval exceeds a critical threshold. It indicates the fact that the imaginary component of tEE can be utilized as a probe of confinement or to signal confinement/deconfinement phase transitions \cite{Afrasiar:2024lsi}. In pseudoentropy it can also detect the chirality of link states associated with topological links \cite{Caputa:2024qkk}. In this work, we add one more crucial property of the imaginary part of the tEE, by showing that it indicates the presence of Fermi surfaces.

In theories with a Lifshitz-like spatial anisotropic symmetry, Lifshitz symmetry or hyperscaling violation, both real and imaginary components of the tEE depend non-trivially on theory parameters that characterize the anisotropy and the breaking of the scaling invariance. Particularly for Lifshitz theories, tEE’s sensitivity to time symmetry breaking contrasts with holographic entanglement entropy, which localizes the time direction and lacks explicit dependence on Lorentz-breaking features at zero temperature.

The non-relativistic theories do not automatically satisfy the Null Energy Condition (NEC) \cite{Dong:2012se,Chu:2019uoh,Hoyos:2021vhl}, which can be thought of as the minimal natural conditions to have a holographic theory that ensures non-repulsive gravity \cite{Wald:106274}, and avoids instabilities and superluminal modes in the scalar correlators of the theory. The satisfaction of the NEC sets a subregion in the parametric space of the parameters that measure the degree of anisotropy $z$ and hyperscaling violation $\th$. Nevertheless, we study the tEE on the whole parametric space of the theory. We provide a classification of the timelike and spacelike surfaces that comprise the tEE, depending on the values of the parameters and we find that there is a correlation between the behaviors of the tEE surfaces and the NEC-compliant parameters. In particular, when the NEC is satisfied, the tEE surfaces in non-relativistic theories have common characteristics with the surfaces of conformal field theories. Additionally, we show that when extra natural conditions are imposed on the tEE, the parametric space of the theory is constrained to be almost identical to the one that NEC and thermodynamic stability conditions of the theory imposed. Our results suggest that the tEE encodes the stability and naturality of a theory.

The real and imaginary parts of the tEE are related to each other through a relation we specify and involve the parameters and the dimension of the theories. This finding is also dictated by  an appropriate dimensional analysis. In the limit of the large dimensions, we show that, irrespective of the theory we are working on, the real and the imaginary parts tend to become equal. We believe that this is a universal property of the tEE.

Moreover, in theories with Lifshitz-like spatial anisotropy, we find that both real and imaginary parts are  heavily dependent on the measurement direction and on the anisotropy degree of the theory, which is common for non-local observables in anisotropic holography \cite{Giataganas:2012zy}.

We show that the tEE also serves as a criterion to identify Fermi surfaces. In particular, they can be defined via the logarithmic behavior of the real part of tEE. Alternatively, they can be defined via a constant imaginary part of the tEE of a certain value we specify that depends on the Lifshitz exponent of the theory. This finding provides an additional meaning for the imaginary part of the tEE.

Whenever we compute the tEE, we accompany it with a computation of the temporal entanglement entropy in the Euclidean signature. The temporal entanglement entropy depends on the theory parameters as well.  After a Wick rotation on the interval, it may be purely imaginary, real or complex depending on the dimensionality of the system, Lifshitz exponent and the exponent of hyperscaling violation.  As we explicitly pointed out, in general the temporal entanglement entropy is a different quantity than the tEE defined above. Nevertheless, the temporal entanglement entropy hints as well at the presence of the Fermi surfaces in the system.

Other relevant recent studies on the tEE have been focused in hyperbolic AdS, dS, in the special case of BTZ black holes and in AdS-Schwarzschild-like black holes, as well as in low dimensional boundary CFTs \cite{Doi:2023zaf,Chu:2023zah,Narayan:2022afv,Li:2022tsv,Narayan:2023zen,Jiang:2023loq,Guo:2024edr,Jena:2024tly};  in confining theories where it has been shown that they can probe confinement \cite{Afrasiar:2024lsi}; and
in the special case of the ($2+1$)-dimensional holographic Lifshitz theories including the temporal entanglement entropy \cite{Grieninger:2023knz,Jena:2024tly,Basak:2023otu} and other related studies including the links with  pseudoentropies  \cite{Doi:2022iyj,Kanda:2023jyi,Caputa:2024gve,He:2024jog,Jiang:2023ffu,Guo:2024lrr,Heller:2024whi}.

The plan of the paper is as follows. In section \ref{sec_tee_temp} we introduce the tEE and the temporal entanglement entropy for generic holographic homogeneous spacetimes with a potential presence of spatial anisotropy. Moreover, we introduce the unnormalized gradient normal vector fields on the spacelike and timelike tEE surfaces in a generic spacetime. In section  \ref{aniso_2d}, we study the tEE and the temporal entanglement entropy of holographic theories with Lifshitz-like spatial anisotropy. In section \ref{sec_hyp}, we discuss theories with hyperscaling violation and the special case of the Lifshitz theories. We discuss in detail the classification of the surfaces with respect to their behavior and how this is related to the stability and NEC of the theory. Moreover, we show that the behavior of the tEE can be used as a criterion of the presence of Fermi surfaces. We finish our manuscript by a short section where we discuss some of our findings and an appendix which contains a discussion of the NEC, to support the main text.

\section{Timelike and Temporal Entanglement Entropy}\label{sec_tee_temp}

\subsection{Timelike Entanglement Entropy}

Let us formulate the setup by considering a generic metric with $d+1$ dimensions as
\begin{equation}\label{genmet}
ds^2_{d+1}=g_{tt}(r)dt^{2}
	+g_{xx}(r)dx_i^2+g_{yy}(r)dy_j^2+g_{rr}(r)dr^{2}~,
\end{equation}
where $r$ is the holographic  direction, $t$ is the time direction, the spatial coordinates $x_i$ and $y_i$ extend respectively along $d_1$ and $d_2$ directions such that $d_1+d_2=d-1$. The boundary of the space can be considered at $r\to 0$ in this section without loss of generality. Let us contemplate the tEE of an infinite strip-like subsystem, $A= -T/2 < t < T/2$ of length $T$ and $L$ in the time direction $t$ and the transverse directions respectively, say at the fixed $x_1=0$ slice on the asymptotic boundary. As usual, $L$ is infinitely large to eliminate the corner contributions. The tEE can be computed by extremizing the area integral
\begin{equation}\label{genSint}
	\mathcal{S}^T=
	\ff{L^{d-2}}{2G_N^{d+1}}	\int_{r_d}^{r_u} dr~~g_{xx}^{\frac{d_1-1}{2}}~g_{yy}^{\frac{d_2}{2}}~\sqrt{g_{rr}+g_{tt}\, t^\prime(r)^2}~,
\end{equation}
where $r_d$ and $r_u$ are the lower and the upper bounds of the integral which we will specify later and depend on the type of spacetime and surface we consider. Note that, the strip-like subsystem can also be considered at a fixed $y_j=0$ slice on the asymptotic boundary. In that case, the metric components $g_{xx}$ and $g_{yy}$ will have powers $\frac{d_1}{2}$ and $\frac{d_2-1}{2}$ respectively in \eq{genSint}. The generalization in these two different directions is straightforward and therefore in this section, we will proceed with a strip-like subsystem at a fixed $x_1=0$ slice. Utilizing \eq{genSint}, the equation of motion reads as
\begin{equation}\label{gentprime}
	t^\prime(r){}^2=\frac{c^2 ~g_{rr}}{ g_{tt}~(g_{tt}~g_{xx}^{d_1-1}g_{yy}^{d_2}  -c^2)}~,
\end{equation}
where $c$ is an integration constant. Here we consider $c^2=sC^2$ where $C^2=-g_{tt}\,g_{xx}^{d_1-1}g_{yy}^{d_2}|_{r=r_0}=:-g_{tt_0}\,g_{xx_0}^{d_1-1}g_{yy_0}^{d_2}>0$ and $s=\pm1$. Where, $s=-1$ corresponds to the equation of the timelike surface, while for $s=1$ we have the equation of the spacelike surface. This will become even more obvious from the generic expression of the gradient normal vector field \eq{pptt} in the later part of this article.  Furthermore, we compute the length of the subsystem in the boundary corresponding to the surface in \eq{gentprime} as,
\begin{equation}\label{gensub}
	T=2
	\int_{r_d}^{r_u} dr~t^\prime(r)~.
\end{equation}
When $s=-1$, the surface in \eq{gentprime} shows a turning point and can be restricted potentially in the region $\infty>r\geq r_0$. This is the timelike surface, which when considered in the conformal AdS Poincare patch, has boundary conditions $t^\prime(r_0)=\infty$ and $t^\prime(\infty)=\pm \infty$. Utilizing \eq{genSint} and \eq{gentprime} with $r_d=r_0$ and $r_u=\infty$, the area of this timelike surface can be computed which yields imaginary value and is free from any divergences as it does not approach the boundary. We will call this quantity $\hat{\cS}^T_{Im}$ in the rest of this article. In terms of the generic metric elements described in \eq{genmet}, $\hat{\cS}^T_{Im}$ can be expressed as
\begin{equation}\label{genSintIm}
	\hat{\cS}^T_{Im}=
	\ff{L^{d-2}}{2G_N^{d+1}}	\int_{r_0}^{r_u} dr~~g_{xx}^{\frac{d_1-1}{2}}~g_{yy}^{\frac{d_2}{2}} \sqrt{\frac{g_{rr}~g_{tt}~g_{xx}^{d_1-1}~g_{yy}^{d_2}}{g_{tt}~g_{xx}^{d_1-1}~g_{yy}^{d_2}-g_{tt_0}~g_{xx_0}^{d_1-1}~g_{yy_0}^{d_2}}}.
\end{equation}
Besides, we can compute the subsystem length corresponding to the timelike surface utilizing \eq{gensub} for $s=-1$ and name it as $T_{Im}$.

For $s=+1$, the surface does not obtain any turning point and it extends from the boundary to the deep IR region. This is the spacelike surface which in the conformal AdS Poincare patch obeys the boundary condition $t^\prime(0)=0$ and $t^\prime(\infty)=\pm \infty$. The area of this spacelike surface $\cS^T_{Re}$ can be computed from \eq{genSint} with limits $r_d=0$ and $r_u=\infty$. Utilizing similar limits, we compute the length of the subsystem $T_{Re}$ in the boundary corresponding to the spacelike surface. Note that, experiencing the infinite proper length from the UV region, $\cS^T_{Re}$ suffers from UV divergences. We remove the divergence by subtracting the area of another solution of the \eq{gentprime} with $t^{\prime}(r)=0$. This solution is described by a pair of straight surfaces extending from the boundary to the deep IR region. The corresponding area integral of these straight disconnected surfaces is given by
\begin{equation}\label{genSintDisc}
	\mathcal{S}^T_{discon}=
	\ff{L^{d-2}}{2G_N^{d+1}}	\int_{0}^{\infty} dr~~g_{xx}^{\frac{d_1-1}{2}}~g_{yy}^{\frac{d_2}{2}} \sqrt{g_{rr}}.
\end{equation}
We will denote the finite part of the spacelike surface as $\hat{\mathcal{S}}^T_{Re}=\mathcal{S}^T_{Re}-\mathcal{S}^T_{discon}$. We can express this area integral in terms of the generic metric described in \eq{genmet} as follows
\begin{equation}\label{genSintRe}
	\hat{\mathcal{S}}^T_{Re}=
	\ff{L^{d-2}}{2G_N^{d+1}}	\int_{0}^{\infty} dr~~g_{xx}^{\frac{d_1-1}{2}}~g_{yy}^{\frac{d_2}{2}} \left(\sqrt{\frac{g_{rr}~g_{tt}~g_{xx}^{d_1-1}~g_{yy}^{d_2}}{g_{tt}~g_{xx}^{d_1-1}~g_{yy}^{d_2}+g_{tt_0}~g_{xx_0}^{d_1-1}~g_{yy_0}^{d_2}}}-\sqrt{g_{rr}}\right).
\end{equation}

\subsection{Temporal Entanglement Entropy}

To consider the temporal entanglement entropy, we introduce in the metric \eq{genmet} in Euclidean signature with the replacement $\tau= i\, t$ where $\tau$ is the Euclidean time direction. In this construction, we study a strip-like subsystem $A=-T/2<\tau<T/2$ in the Euclidean time direction $\tau$ at a constant $x_1$ slice on the boundary $r=0$. In the transverse directions, the length of the $A$ is $L$, where $L$ is very large. The temporal entanglement entropy may then be obtained from the following minimized area integral
\begin{equation}\label{genSint_tau}
	\mathcal{S}^{\tau}=
	\ff{L^{d-2}}{2G_N^{d+1}}	\int_{r_d}^{r_u} dr~g_{xx}^{\frac{d_1-1}{2}}~g_{yy}^{\frac{d_2}{2}}~\sqrt{g_{rr}+g_{\tau\tau}\, \tau^\prime(r)^2}~.
\end{equation}
From the above expression, we can obtain the equation of motion
\begin{equation}\label{gen_tau_prime}
	\tau^\prime(r){}^2=\frac{\tilde{c}^2 ~g_{rr}}{ g_{\tau\tau}~(g_{\tau\tau}~g_{xx}^{d_1-1}g_{yy}^{d_2}-\tilde{c}^2)}~,
\end{equation}
where $\tilde{c}$ is an integration constant.
For the case where we have a connected surface satisfying the above equation of motion, contains a turning point at $r=r_0$. The temporal entanglement entropy for this surface is given by the integral described in \eq{genSint_tau} with $r_d=0$ and $r_u=r_0$ and the integration constant is computed from the turning point $r_0$ as $\tilde{c}^2=g_{\tau\tau}g_{xx}^{d_1-1}g_{yy}^{d_2}|_{r=r_0}$ where $\tau^{\prime}(r_0)=\infty$. The area integral contains UV divergences at the boundary $r\rightarrow 0$ which is the same as the divergence of two straight disconnected surfaces located at $t=\pm \frac{T}{2}$ and stretched between the boundary $r=0$ and the deep IR region $r=\infty$. These surfaces satisfy the equation of motion $\tau^{\prime}(r)=0$ and thus the area integral becomes
\begin{equation}\label{genSint_tau_disc}
    \mathcal{S}_{discon}^{\tau}=
    \ff{L^{d-2}}{2G_N^{d+1}}	\int_{r_d}^{r_u} dr~g_{xx}^{\frac{d_1-1}{2}}~g_{yy}^{\frac{d_2}{2}}~\sqrt{g_{rr}}~,
\end{equation}
with $r_d=0$ and $r_u=\infty$. The subsystem length can be obtained by integrating \eq{gen_tau_prime} with respect to the holographic direction $r$ as follows
\begin{align}
    T= \int_{r_d}^{r_u} dr~\frac{\tilde{c}^2 ~g_{rr}}{ g_{\tau\tau}~(g_{\tau\tau}~g_{xx}^{d_1-1}g_{yy}^{d_2}-\tilde{c}^2)}~,
\end{align}
with $r_d=0$ and $r_u=r_0$.

\subsection{Timelike Entanglement Entropy Surface Properties}\label{cov}

In this subsection, we introduce the spacelike and timelike surfaces and discuss their properties. We first consider the tEE hypersurface consisting of the spacelike or timelike surfaces as,
\begin{align}\label{spacelike_timelike}
    \Sigma_{Im} := t-\int_{r_d}^{r_u} dr~t_{Im}^\prime(r)=0~\qquad \Sigma_{Re} := t-\int_{r_d}^{r_u} dr~t_{Re}^\prime(r)=0~,
\end{align}
where the $t(r)$ is the solution of \eq{gentprime} which is different for the spacelike and the timelike surfaces. The subindex $Im$ and $Re$ refer to the timelike and spacelike surfaces which have imaginary and real area contributions respectively.
Once the hypersurface $\Sigma$ is determined, the tangent and transverse vector to $\Sigma$ can be defined.

The tangent and transverse vector fields of the hypersurfaces can be normalized and rescaled by non-zero and non-diverging functions. Moreover, the expressions of the vectors change in different coordinate systems and parametrizations, while the areas remain invariant. Here we have chosen the static gauge to parametrize the minimal surface and we are working with this parametrization. We  consider the unnormalized gradient normal vector field $g^{\a\b}\pp_\b \S$ satisfying the equation
\be \la{normalv}
\nabla_\a\pp_\b\S-\nabla_\b\pp_\a\S=0~.
\ee
Notice that generic rescaled normal vector fields to the hypersurface $\S$ do not satisfy  \eq{normalv}. Moreover, the gradient vector field is not of constant length since the hypersurface is not a geodesic. Furthermore, for the tangent vector $\cP$ we choose a simple variation of the parametrization, related directly to the normal vector field. We introduce
\begin{align}
\cT_\a=\partial_\a\Sigma=\left(1,0,\emptyset_{d_1-1},\emptyset_{d_2},-t^\prime(r)\right)~,\qquad
\cP^\a=\left(t^\prime(r),0,\emptyset_{d_1-1},\emptyset_{d_2},1\right)~,
~\label{zeroset}
\end{align}
such that $\cP^\a\cT_\a=0$. Each $\cP^\a_{\prt{Re,Im}}$ corresponds to each of $t_{\prt{Re,Im}}^\prime(r)$ and $\emptyset_q:=(0,0,\cdot\cdot\cdot,0)$ possessing  $q$ components respectively.

A  gradient normal $\tilde\cT^\a$ vector at a constant $r=r_c$ hypersurface $\S_r:=r-r_c$, and the tangent $\tilde\cP^\a$ to $\S_r$ read from
\begin{align}
    \tilde\cT_\a:=(0,0,\emptyset_{d_1-1},\emptyset_{d_2},1)~,\qquad  \tilde\cP^\a:=(1,0,\emptyset_{d_1-1},\emptyset_{d_2},0)~.
\label{ortho1}
\end{align}
 as $\tilde\cP^\a\tilde\cT_\a=0$. We define for the vectors   discussed so far the following quantities
\begin{align}\la{pptt}
|\cT|^2&=\cT_\a~g^{\a\b}~\cT_\b=\frac{g_{xx}^{d_1-1}g_{yy}^{d_2}}{~ g_{tt}~g_{xx}^{d_1-1}g_{yy}^{d_2}  -s~ C^2 } ~,\qquad |\cP|^2=\cP^\a~g_{\a\b}~\cP^\b=g_{rr}~g_{tt}~|\cT|^2~,~\\
|\tilde\cP|^2&=\tilde\cP^\a~g_{\a\b}~\tilde\cP^\b=g_{tt}~,\qquad
|\tilde\cT|^2=\tilde\cT_\a~g^{\a\b}~\tilde\cT_\b=g^{rr}~.
\end{align}
Notice that $ |\cP|^2$ and $ |\cT|^2$ have opposite signs as expected. A timelike surface has positive $|\cT|^2$ for the transverse vector, while negative $|\cP|^2$ for the tangent vector. The equation \eq{pptt} for $|\cT|^2$ provides an additional implication that has not been noticed before. It essentially shows how the timelike and spacelike surfaces behave. In any holographic theory, for timelike and spacelike surfaces to asymptotically end at the same location, either the norm $|\cT|^2$ must be zero at this point, effectively making them appear null or the subvolume formed by the time and spatial directions excluding the direction in which the strip is localized, specifically $g_{tt}~g_{xx}^{d_1-1}g_{yy}^{d_2}$ in this case, must vanish, leading to $|\cT|^2 \sim -s^{-1} g_{xx}^{d_1-1}g_{yy}^{d_2}C^{-2}$. These are the only allowed options.

Utilizing the above expressions, we introduce two more quantities, which can be thought as orthogonality measures of the tEE hypersurfaces with planes of constant $r$ including the boundary of the theory:
\bea\la{ii}
&&|I_1|^2:=\prt{\frac{\cP^\a~g_{\a\b}~\tilde\cP^\b}{|\cP||\tilde\cP|}}^2=\prt{\frac{\cT_\a~g^{\a\b}~\tilde\cT_\b}{|\cT||\tilde\cT|}}^2=\frac{s~C^2}{g_{tt}\, g_{xx}^{d_1-1} g_{yy}^{d_2}}~,
\\
&&|I_2|^2:=\prt{\frac{\cP^\a~g_{\a\b}~\tilde\cT^\b}{|\cP||\tilde\cT|}}^2=\prt{\frac{\cT^\a~g_{\a\b}~\tilde\cP^\b}{|\cT||\tilde\cP|}}^2=1-|I_1|^2~.\la{ii2}
\eea

All the generic formulas presented so far are valid for an $x_1=0$ localization of the boundary. However, all these relations can be generalized for the strip localized along a $y_j$ direction following the procedure described below equation \eq{genSint}. In the next section, we consider more systematically, the properties of the tEE surfaces.

\section{Timelike Entanglement Entropy in Anisotropic Lifshitz-like Theories} \label{aniso_2d}
Let us first study theories with a Lifshitz-like anisotropic symmetry. These theories are scale invariant, they have a spatial anisotropy and the background has the following form,
\begin{align}\label{aniso}
    ds_{d+1}^2=-\frac{dt^2}{r^{2 z}}+\frac{dx_i^2}{r^{2 z}}+\frac{dy^2+dr^2}{r^2}~,  %\qquad  i=(2,3,...(d-2))~,
\end{align}
with $i=(1,2,...d_1)$ and $d_1=(d-2)\ge0$ since $d \geq 2$ and $d_2=1$. The $d_2$ dimensions of $y$ can be chosen arbitrarily, here without much loss of generality we set them equal to one and we discuss when this choice matters and its implications below in detail.  %\textcolor{red}{(**)}.
The theory remains invariant under a rescaling of the space-time coordinates
\begin{align}
    t \rightarrow \lambda^z t~, \quad x \rightarrow \lambda^z x~, \quad x_i \rightarrow \lambda^z x_i~, \quad y \rightarrow \lambda\, y~,
\end{align}
with $z$ is a measure of the degree of Lorentz symmetry violation and anisotropy. Notice that this type of theory has been obtained with a fixed exponent $z=3/2$ by considering the backreaction of space-dependent axion fields in the IIB supergravity \cite{Azeyanagi:2009pr}, as well with an arbitrary exponent $z$ in generic Einstein-Axion-Dilaton systems \cite{Giataganas:2017koz}.

There is a special limit to obtain the Lifshitz theory from the anisotropic theory, by  "eliminating" the $d_1$  $x-$spatial dimensions to recover the usual Lifshitz theories. In fact, observables that localize along the volume $x_i$ would behave as being in a Lifshitz theory. In addition, observables that localize along the $y_i$-volume would behave as being in an AdS conformal theory. This is one of the cases we present below.

\subsection{Timelike Entanglement Entropy localized on \texorpdfstring{$y$}~ direction}\label{aniso_timelike}

We choose a constant $y$ slice corresponding to the metric described in \eq{aniso} and analyze the different types of surfaces. We subsequently compute the tEE for this setup. The results of this warm-up computation can be obtained by coordinate transformation of the AdS spacetime, as we show explicitly below.

\subsubsection{Timelike Entanglement Entropy Surface Properties}\label{aniso_cov}

The equations of motion \eq{gentprime} for the spacelike and timelike parts of the extremal surfaces contributing to the tEE are described respectively as
\begin{align}\label{EOM_y_slice_timelike0} t^{\prime}=\frac{r^{z-1}}{\sqrt{1+s\left(\frac{r_0}{r}\right){}^{2 (d-1) z}}}~.
\end{align}
We notice the two types of surfaces based on the sign of $z$. For positive $z$ the timelike surface extends from its turning point asymptotically
to the bulk where we have
$     t_{Im}^{\prime}|_{r\rightarrow \infty}=t_{Re}^{\prime}|_{r\rightarrow \infty}= r^{z-1}~.$
For these types of hypersurfaces, and $z\ge1$ we have $t_{Re}^{\prime}$ to vanish at the boundary.
When $z$ is negative, the timelike surface extends from its turning point to $r=0$ asymptotically, but at the same time, the metric elements $x$ diverge now at the opposite direction $r\rightarrow \infty$.

The corresponding measures of the transverse and parallel vectors are computed as
\bea\label{cov_aniso_y}
&&|\cT|^2=-\frac{r^{2 z}}{1+s\left(\frac{r}{r_0}\right)^{2 (d-1) z}},\qquad |\cP|^2=-r^{-2\prt{z+1}}|\cT|^2~~\\
&&|I_{1}|^2=  s\left(\frac{r}{r_0}\right)^{2(d-1) z},\qquad
|I_{2}|^2= 1-s\prt{\frac{r}{r_0}}^{2 (d-1) z}
\eea
where we remind that $s:=\pm1$, with the positive sign corresponding to the spacelike surface and the negative to the timelike surface. Immediately we notice that for the spacelike surface  which extends from the boundary to the deep IR always $|\cP|^2=-r^{-2(z+1)}|\cT|^2>0$,
while for the timelike surface $|\cP|^2=-r^{-2(z+1)}|\cT|^2<0$, as long as
\bea
&&\mathbf{A:}~ z>0~,\qquad r_{Im}\ge r_0~, \qquad \mbox{or},\qquad \mathbf{B:}~ z<0~,\qquad r_{Im}\le r_0~,\la{caseB0}
\eea
where $r_0$ is the turning point of the surface.

The behavior of the surfaces of the tEE is obviously coordinate system dependent, although it is highly instructive. Let us discuss firstly case A, where the boundary of the theory is at $r=0$ while the timelike surface extends from the turning point $r_0$ to the deep IR, therefore $r_{Im}\ge r_0$ as long as $z>0~,$ which can also be extracted from the surface equations \eq{EOM_y_slice_timelike0} by requiring them to have real solutions.
Moreover, $|I_{1}|$ is vanishing in the UV limit $r\rightarrow 0$ for the spacelike surface, and $|I_{2}|$ is non-vanishing.

Nevertheless, the asymptotic behavior between the surfaces differ for $z\ge1$, $d^{-1}\le z<1$ and $0\le z\le d^{-1}$.  For $z\ge1$ the spacelike surface approaches the boundary perpendicular, while the timelike surface extends from its bulk turning point to the deep IR asymptotically with diverging $t'$. This behavior resembles closely the conformal tEE surfaces $(z=1)$. In the case $d^{-1}\le z\le1$ both in the boundary and the deep IR region, the surfaces have vanishing $t'$.

In the case $0\le z\le d^{-1}$ the spacelike tEE surface has a diverging $t'$ at the boundary and a vanishing one in the IR region. Also, the timelike surface approaches the IR region with vanishing $t^{\prime}$.

The surfaces of type B have a common behavior, the timelike surface exists for $r<r_0$ and approaches asymptotically the $r=0$ with diverging $t'$.  The spacelike surfaces also have diverging $t^{\prime}$ at $r=0$ and it approaches $r\rightarrow \infty$ perpendicularly with vanishing $t^{\prime}$.  This is natural since only one $y$ direction is localized in the computation, the observable sees essentially the volume $x_i$ with its metric element diverging at $r\rightarrow \infty$, where the observable effectively sees it as the boundary of the theory. We point out that this is a special case since we have only one $(d_2=1)$ localized dimension $y$ in the anisotropic space, and in the next section \ref{x_dire} with a localization along the $x$ directions, we will have a different outcome.

A careful dimensional analysis gives for the tEE
\be \la{teely}
\hat{\mathcal{S}}_{Im}\sim \hat{\mathcal{S}}_{Re}\sim T^{-\prt{d_1+\frac{d_2-1}{z}}}\sim T^{-\prt{d-2}},
\ee
which decreases with increasing time interval $T$ as expected. Notice the speciality of the choice $d_2=1$, as this is the only case where the tEE can be independent of the scaling exponent $z$.

The classification of the behavior of the extremal surfaces with $z$-parameters, is coordinate and parametrization dependent, although in our analysis as we will see in the manuscript, is highly instructive. Nevertheless, applying it for the $y$-localized strip we would obtain a constrain on $z$, while $z$ here can be absorbed in the radius of the induced localized spacetime and in this case the constrain has no particular physical value.

\subsubsection{Equivalence with the Conformal Theory's Timelike Entanglement Entropy}\label{conformal}

Before we proceed to presenting the explicit computations, let us further discuss the tEE for intervals localized along the $y$-direction, in a homogeneous static, and therefore diagonal metric, with one $(d_2=1)$ anisotropic direction.
By doing the coordinate transformation
\be
r=\tilde{r}^{1/z}~,\quad \mbox{and}\quad \prt{t,x}=\prt{\frac{\tilde{t}}{z},\frac{\tilde{x}}{z}}~,
\ee
we note that we are effectively working in AdS localized spacetime with a rescaled radius and one dimension down. The localized $y$-metric \eq{aniso} under this coordinate transformation  becomes
\begin{align}\label{anisoloc}
    d\tilde{s}_{d+1}^2=\frac{1}{z^2}\frac{1}{\tilde{r}^2}\prt{-d\tilde{t}^2+d\tilde{x}_i^2+d\tilde{r}^2}~.
\end{align}
Therefore the tEE in this case should be equal to
\be \la{ads1}
\hat{\mathcal{S}}\sim \frac{1}{z^{d-1}} \hat{\mathcal{S}}_{AdS, R=1}\sim \frac{1}{z^{d-1}}   T^{-\prt{d-2}}~,
\ee
where $\hat{\mathcal{S}}_{AdS, R=1}$ is the tEE in the AdS spacetime of radius $R=1$. A rescaling of the radius $R$ would absorb the $z$-dependence. The scaling power in \eq{ads1} agrees with \eq{teely} as expected.

\subsubsection{y-localized Timelike Entanglement Entropy}

We now proceed to explicitly compute the subsystem lengths corresponding to the surfaces $\Sigma_{Im}$ and $\Sigma_{Re}$ by integrating respectively $2 t_{Im}^{\prime}$ and $ 2t_{Re}^{\prime}$ described in \eq{EOM_y_slice_timelike0} with respect to the bulk direction $r$. As we have described in the previous section, the tEE is proportional to the tEE of AdS  \eq{ads1} with a rescaled metric and in principle, we could do the computation in the AdS metric \eq{anisoloc} which to our knowledge is also not presented anywhere so far. Nevertheless, since there is no extra complication we proceed to the computation in the equivalent $y$-localized metric \eq{aniso}.

Without loss of generality let us consider $z>0$, and we obtain the subsystem lengths $T_{Im}$ and $T_{Re}$ from \eq{gensub} as follows,
\begin{align}
     T_{Im}&=\frac{2\, \epsilon_1 ^{-z}}{z}- \frac{2\, r_0^z}{z}\frac{\sqrt{\pi }\, \Gamma \left(1-\alpha\right)}{\Gamma \left(\frac{1}{2}-\alpha\right)}~,\label{Tim_yslice}\\
     T_{Re}&=\frac{2\, \epsilon_1 ^{-z}}{z}- \frac{2\, r_0^z}{z} \frac{\Gamma \left(1-\alpha\right) \Gamma \left(\frac{1}{2}+\alpha\right)}{\sqrt{\pi }}~,\label{Tre_yslice}
\end{align}
where $\alpha:=\prt{2 (d-1)}^{-1}$ is defined for presentation purposes and depends on the inverse of the number of dimensions. In the above expressions, $\epsilon_1 \ll 1$ is introduced at the IR region to keep track of the infinite lengths $T_{Im}$ and $T_{Re}$. As in \cite{Afrasiar:2024lsi}, at the IR region $r\rightarrow \infty$, the extremal surfaces contributing to the real and imaginary parts of the tEE satisfies $t^{\prime}_{Re}\left(\infty\right)=t^{\prime}_{Im}\left(\infty\right)$ in the Poincar\'e coordinate system with the given parametrization, then the total subsystem length at the boundary $r \rightarrow 0$ is
\begin{align}
    T = T_{Im}-T_{Re}
    = \frac{2\, r_0^z}{z} \frac{\sqrt{\pi }\,  \Gamma (1-\alpha )\left(\sec (\pi  \alpha )-1\right)}{\Gamma \left(\frac{1}{2}-\alpha \right)}~.
\end{align}
From the above expression, we can obtain the turning point $r_0$ in terms of the total subsystem length $T$ as
\begin{align}\label{r0_y_slice}
    r_0= \left(\frac{z\, \Gamma \left(\frac{1}{2}-\alpha \right)}{2 \sqrt{\pi }\, \Gamma (1-\alpha )\left(\sec (\pi  \alpha )-1\right)}\,T\right)^{1/z}~.
\end{align}
The imaginary part and the finite real part of the tEE can be obtained respectively from \eq{genSintIm}  and \eq{genSintRe} as
\begin{align}
    \frac{4G_N^{(d+1)}}{L^{d-2}}\hat{\mathcal{S}}_{Im}^T
    =-i\, \ff{2r_0^{\left(1-\frac{1}{2 \alpha } \right)  z}}{z}\ff{ \alpha\Gamma \left(1-\alpha\right) \Gamma \left(-\frac{1}{2}+\alpha\right)\cos \left(\pi \alpha\right)}{\sqrt{\pi\, }}~.
\end{align}
\begin{align}
    \frac{4G_N^{(d+1)}}{L^{d-2}}\hat{S}_{Re}^T = \frac{4G_N^{(d+1)}}{L^{d-2}}\left(\mathcal{S}^T_{Re}-\mathcal{S}^T_{discon}\right)
    &=\frac{2r_0^{\left(1-\frac{1}{2 \alpha } \right) z}}{z} \frac{ \alpha\Gamma \left(1-\alpha\right) \Gamma \left(-\frac{1}{2}+\alpha\right)}{\sqrt{\pi }}~.
\end{align}
Utilizing \eq{r0_y_slice}, the above expressions can be written in terms of the total subsystem length. For the imaginary part we get
\be\label{Sim_yslice}
    \frac{4G_N^{(d+1)}z^{d-1}}{L^{d-2}}\hat{\mathcal{S}}_{Im}= i \frac{2^{d-1} \pi^{\frac{d-1}{2}}}{d-2} \left(\sec \left(\frac{\pi }{2-2 d}\right)-1\right)^{d-2} \left(\frac{\Gamma \left(1+\frac{1}{2-2 d}\right)}{\Gamma \left(\frac{d-2}{2 (d-1)}\right)}\right)^{d-1}\left(\frac{1}{T}\right)^{d-2},
\ee
and for the real
\be \la{Sre_yslice}
\hat{\mathcal{S}}_{Re}=i\sec \left(\frac{\pi }{2-2 d}\right)\hat{\mathcal{S}}_{Im}.
\ee
Notice that the $z^{d-1}$ dependence appears exclusively in the left hand side of \eq{Sim_yslice} and represents the rescaling of the radius $R=1$ of the AdS space to $R=1/z$. Therefore the real and imaginary parts of tEE presented in \eq{Sim_yslice} and \eq{Sre_yslice} are identical to the tEE in conformal AdS spacetime. While the imaginary and real areas have norms that differ by a $d$-dependent factor $\sec \prtt{\frac{\pi}{2}\prt{d-1}^{-1}}$. In the large $d$ limit, we note $|\hat{S}_{Im}^T|\simeq|\hat{S}_{Re}^T|$, suggesting that in the large dimensions the tEE exhibits significant simplifications, making it worthwhile to study further, as in the case of the entanglement entropy in \cite{Giataganas:2021jbj}.

\subsection{Timelike Entanglement Entropy localized on \texorpdfstring{$x_1$}~ direction} \la{x_dire}

The anisotropic theories have distinct tEE depending upon the direction we consider the entangling degrees of freedom. To present a complete analysis we need to consider a constant slice in the $x_1$ direction of the theory described by \eq{aniso}.

\subsubsection{Timelike Entanglement Entropy Surface Properties}\label{aniso_cov_x1}

\begin{figure}[t]
\begin{minipage}[t]{0.5\textwidth}
	\begin{flushleft}
		\centering{\includegraphics[width=80mm]{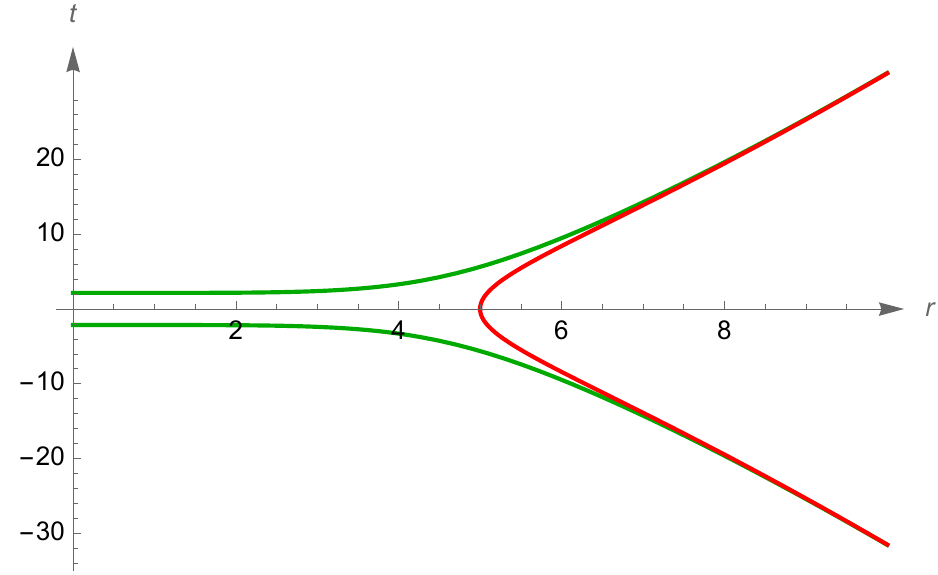}}
		\caption{The tEE surfaces for $z>1$ corresponding to  $A_2$ of \eq{case2A}. The boundary of the theory is at $r\rightarrow 0$. The equation of motion for the spacelike hypersurfaces $\Sigma_{Re}$ follows the boundary conditions $t^{\prime}_{Re}|_{r\rightarrow 0}=0$ and $t^{\prime}_{Re}|_{r\rightarrow \infty}=\infty$, whereas for the timelike surface $\Sigma_{Im}$, the boundary conditions are given by $t^{\prime}_{Im}|_{r\rightarrow r_0}=\infty$ and $t^{\prime}_{Im}|_{r\rightarrow \infty}=\infty$. We plot the timelike surfaces with red color and the spacelike ones with green. Here we choose $d=4$ and the turning point of the timelike surface at $r_0=5$. These choices are same for the neighboring plots of this section. Moreover, here $z=3/2$.}\label{fig:Case1_x1}
	\end{flushleft}
\end{minipage}
\hspace{0.3cm}
\begin{minipage}[t]{0.5\textwidth}
	\begin{flushleft}
		\centering{\includegraphics[width=80mm ]{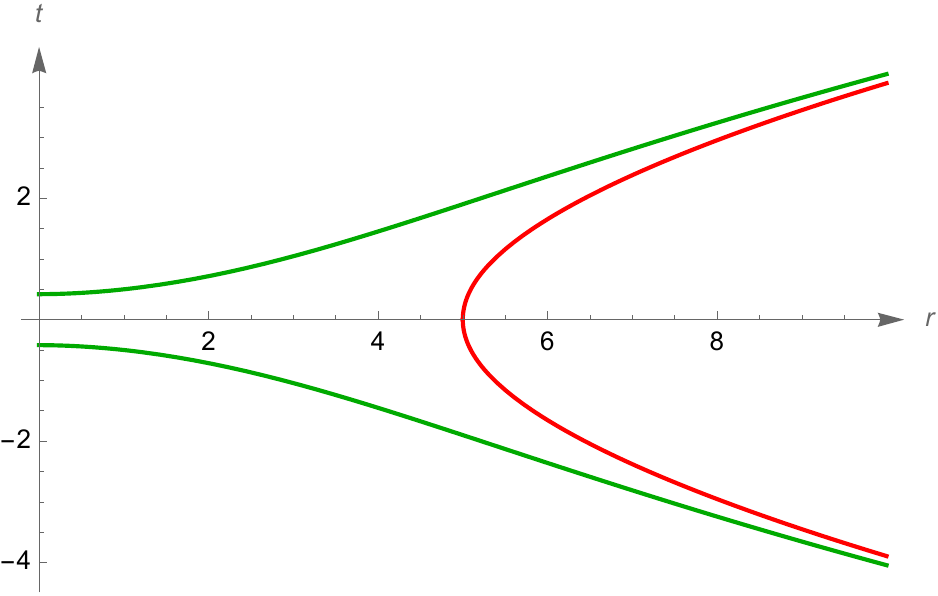}}
		\caption{The surfaces for $0<z<1$ which is discussed in $A_1$ of \eq{case2A}. The boundary of the theory for the range of parameters $A_0$ and case $A_1$ is at $r=0$. The equation of motion for $\Sigma_{Re}$ follows the boundary conditions $t^{\prime}_{Re}|_{r\rightarrow 0}=0=t^{\prime}_{Re}|_{r\rightarrow \infty}$, whereas for $\Sigma_{Im}$ the boundary conditions are given by $t^{\prime}_{Im}|_{r\rightarrow r_0}=\infty$ and $t^{\prime}_{Im}|_{r\rightarrow \infty}=0$. Here we consider $z=3/10$.}
		\label{fig:Case2_x1}
	\end{flushleft}
\end{minipage}
\end{figure}

\begin{figure}[t]
	\begin{minipage}[t]{0.5\textwidth}
		\begin{flushleft}
			\centering{\includegraphics[width=80mm ]{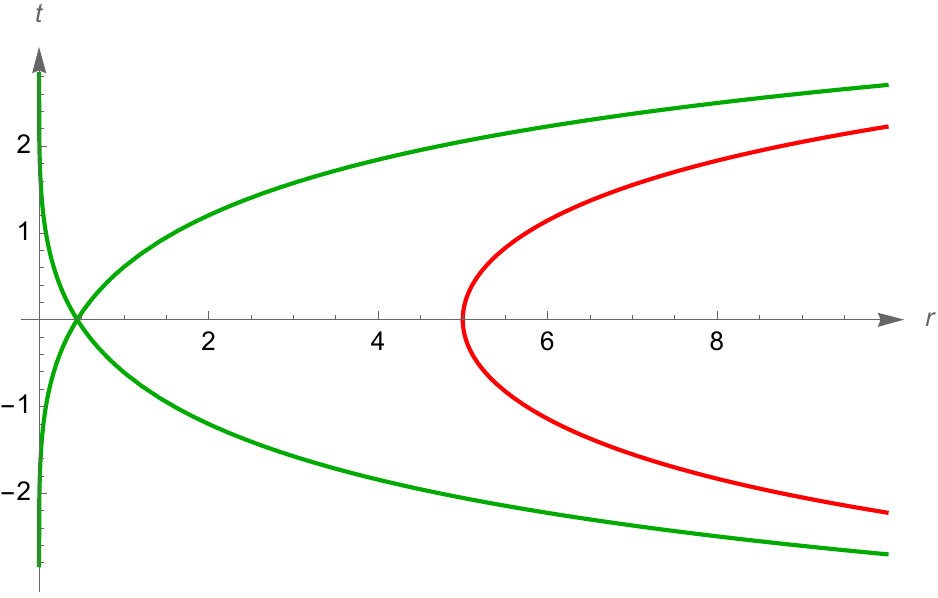}}
			\caption{The  surfaces for $-(d-2)^{-1}<z<0$ which is the $A_0$ of \eq{case2A}. The   $g_{xx}$ and $g_{yy}^{-1}$ diverge  at $r\rightarrow \infty$. The equation of motion for $\Sigma_{Re}$ satisfies $t^{\prime}_{Re}|_{r\rightarrow 0}=\infty$ and $t^{\prime}_{Re}|_{r\rightarrow \infty}=0$, whereas for $\Sigma_{Im}$ we have $t^{\prime}_{Im}|_{r\rightarrow r_0}=\infty$ and $t^{\prime}_{Im}|_{r\rightarrow \infty}=0$. Here we have considered  $z=-1/4$.}
			\label{fig:Case3_x1}
		\end{flushleft}
	\end{minipage}
 	\hspace{0.3cm}
 \begin{minipage}[t]{0.5\textwidth}
		\begin{flushleft}
			\centering{\includegraphics[width=80mm]{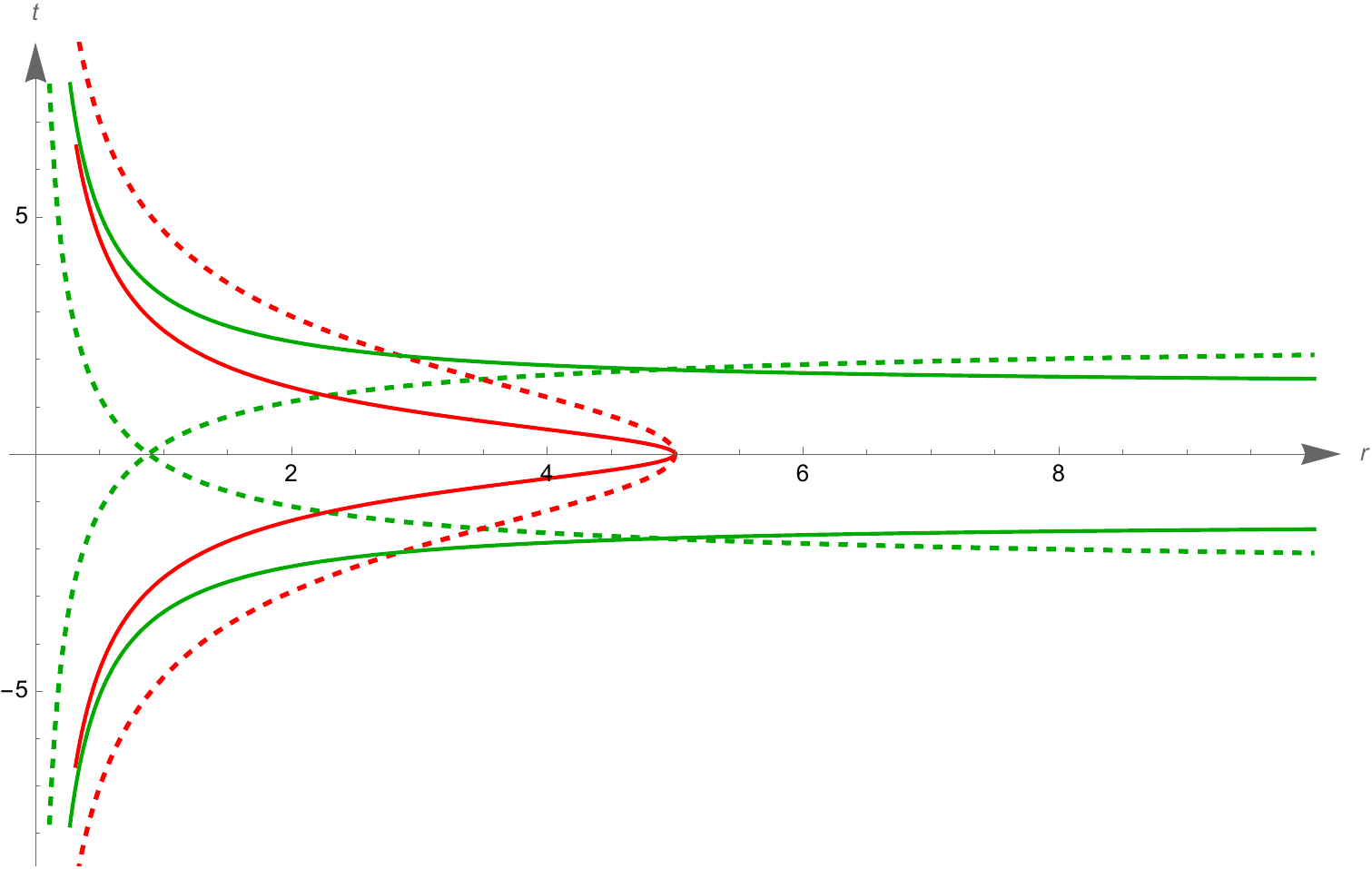}}
			\caption{The hypersurfaces for $z<-(d-2)^{-1}$ described by $B$ of \eq{case2B}.   $g_{xx}$ and $g_{yy}^{-1}$ diverge  at $r\rightarrow \infty$. There are two discrete sets of surfaces for $z=-4/5$ with solid lines and and $z=-3/5$ plotted with dashed. For both the cases, $\Sigma_{Re}$ follow the same boundary conditions given by $t^{\prime}_{Re}|_{r\rightarrow 0}=\infty$ and $t^{\prime}_{Re}|_{r\rightarrow \infty}=0$, while $\Sigma_{Im}$ satisfies $t^{\prime}_{Im}|_{r\rightarrow r_0}=\infty$ and $t^{\prime}_{Im}|_{r\rightarrow 0}=\infty$.}  \label{fig:Case4_x1}
		\end{flushleft}
	\end{minipage}

\end{figure}

In this subsection, we work as in the previous subsection \ref{aniso_cov} to analyze the behaviors of the tEE surfaces with a constant $x_1$ slice. The equations of motion for the surfaces lead to
\begin{align}\label{EOM_x1_slice_timelike}
    t^{\prime}=\frac{r^{z-1} }{\sqrt{1+s\left(\frac{r_0}{r}\right){}^{2 (d-2) z+2}}}~,
\end{align}
where $s=-1$ is the equation for the timelike surface and $s=1$ is for the spacelike ones. When  $(d-2)z>-1$, which includes negative values of $z$,  the timelike surface extends from the turning point to the $r\rightarrow \infty$ where $t_{Im}^{\prime}|_{r\rightarrow \infty}=t_{Re}^{\prime}|_{r\rightarrow \infty}=r^{z-1}~.$
Additionally, we have a second type of surface for $(d-2)z<-1$ where the timelike  surfaces extends from their turning point to $r\rightarrow 0$. We classify the surfaces more systematically below. The corresponding vector quantities read
\bea\label{cov_aniso_x1}
&&|\cT|^2=-\frac{r^{2 z}}{1+s\left(\frac{r}{r_0}\right)^{2 (d-2) z+2}}~,\qquad  |\cP|^2=-r^{-2\prt{z+1}}|\cT|^2\\
&&|I_1|^2=  -s\left(\frac{r}{r_0}\right)^{2(d-2) z+2}~,\qquad
    |I_2|^2= 1-|I_1|^2~.\la{Ix}
\eea
where for timelike surfaces $|\cP|^2=-r^{-2\prt{z+1}} |\cT|^2<0$.
We classify the tEE surfaces by the sign of the power in $I_1$ or equivalently by examining the dominance in the terms of the denominator of the $t'$ described in \eq{EOM_x1_slice_timelike}. In  total, we can split the parametric space as
\bea\la{case2A}
&&\mathbf{A:}~  z\ge-\prt{d-2}^{-1}~\Rightarrow \mathbf{A_0:} -\prt{d-2}^{-1}\le z<0~, \quad\mathbf{A_1:} 0\le z<1~,\quad \mathbf{A_2:} z\ge1~,\qquad\qquad
\\\la{case2B}
&&\mathbf{B:}~ z\le-\prt{d-2}^{-1}~,
\eea
Let us start by discussing the surfaces $A$. For $A_2:z\ge1$, all the metric elements diverge at $r=0$ which is the boundary of the theory in this case. The spacelike surface approaches the boundary perpendicular, while the timelike surface extends from its bulk turning point to the deep IR asymptotically with a non-zero $t'$. A representative surface is plotted in Figure \ref{fig:Case1_x1}. We consider this set of surfaces as having the desirable properties, as they resemble closely the conformal tEE surfaces. Therefore, one could accept the parametric regime $z\ge1$ as the most natural one in the theory. In the case $A_1$, the boundary of the theory remains at $r\rightarrow 0$, the spacelike and timelike surfaces yield similar behaviors as $A_2$ with the difference that both spacelike and timelike surfaces in the deep  bulk tend to have vanishing $t'$. These surfaces have different behavior in the deep IR compared to the conformal ones, but their boundary conditions are sensible. Nevertheless, the regime of Lifshitz exponents  that correspond to, is outside the NEC. A representative surface is plotted in Figure \ref{fig:Case2_x1}. In the case $A_0$ the $x-$metric elements blow up at $r\rightarrow\infty$ while the $g_{yy}$ at the opposite limit. The spacelike surfaces
act as seeing the boundary at infinity where $t'\rightarrow 0$. A representative surface is plotted in Figure \ref{fig:Case3_x1}.
For the case $B$, both spacelike and timelike surfaces act as seeing the boundary at $r\sim \infty$ where now the timelike surface extends from $r_0$ to $\infty$, a set of representative surfaces appear in Figure \ref{fig:Case4_x1}.  Here also the metric elements for $x-$directions blow up at $r\rightarrow \infty$ whereas the y-metric element diverges at $r=0$.

As a side comment, we notice that when $z\ge1$, the norm of transverse vector $|\cT|$ to the tEE spacelike hypersurface vanishes faster compared to the divergence of the tangent vector $|\cP|$  at the boundary. Equivalently,  the norm of $|\cT|$ blows up faster than $|\cP|$ vanishes at the IR for both spacelike and timelike surfaces. Moreover, this condition sets the minimum rate of vanishing $I_1$ for the surfaces at the boundary: $|I_{1}|$ to vanish faster or with the same rate with the corresponding scaling of $|I_{2}|_{\text{AdS}}$ in the UV limit.  Furthermore, this same condition is equivalent of having the rate of the transverse vectors $\cT$ in the non-relativistic conformal field theory diverging slower or with equal rate to the ones in the conformal theory. Moreover, we point out that $I_2$ does vanish for the timelike surface at $r=r_0$ as expected.

In the next section, we will compute the tEE analytically. Nevertheless, utilizing a dimensional analysis the scaling behavior of tEE can be obtained as
\be
\hat{\mathcal{S}}_{Im}\sim \hat{\mathcal{S}}_{Re}\sim T^{-\prt{(d_1-1)+\frac{d_2}{z}}}\sim T^{-\prt{d-3+\frac{1}{z}}}.
\ee
The above relation ensures that the tEE decreases with increasing time for the region $A_2:z\ge 1$ in \eq{case2A} with $d\ge 3$. In summary, the tEE surfaces that have the same properties as the tEE surfaces in conformal theory are the ones for $z\ge1$ which reproduces the NEC and ensures a natural behavior of the tEE.

\subsubsection{Computation of the Timelike Entanglement Entropy}
Let us first compute the subsystem lengths corresponding to the spacelike and timelike surfaces for the metric described in \eq{aniso} at constant $x_1$ slice. We can obtain the subsystems $T_{Im}$ and $T_{Re}$ by integrating the expressions in \eq{EOM_x1_slice_timelike} multiplied by a factor of two to obtain the full intervals as
\begin{align}
    T_{Im}&=\frac{2\, \epsilon_1 ^{-z}}{z}- \frac{2\, r_0^z}{z}\frac{\sqrt{\pi }\, \Gamma \left(1-\beta\right)}{\Gamma \left(\frac{1}{2}-\beta\right)}~,\label{Tim_x1slice}\\
    T_{Re}&=\frac{2\, \epsilon_1 ^{-z}}{z}- \frac{2\, r_0^z}{z} \frac{\Gamma \left(\frac{1}{2}+\beta\right) \Gamma \left(1-\beta\right)}{\sqrt{\pi }}~,\label{Tre_x1slice}
\end{align}
where $\beta:=\frac{z}{2 (d-2) z+2}$, a function of $z$ and $d$, is a new parameter defined for presentation reasons to maintain the expression more compact. $\epsilon_1$ is introduced to keep track of the infinities which cancel in \eq{r0_x1_slice0}.  The total subsystem length can now be expressed in terms of the turning point $r_0$ as
\begin{align}\la{r0_x1_slice0}
    T = T_{Im}-T_{Re}
    = \frac{2\, r_0^z}{z} \frac{\sqrt{\pi }  \Gamma (1-\beta )\left(\sec (\pi  \beta )-1\right)}{\Gamma \left(\frac{1}{2}-\beta \right)}~.
\end{align}
Then the imaginary part and the real part of the tEE can be obtained respectively from \eq{genSintIm}  and \eq{genSintRe} as
\begin{align}
    \frac{4G_N^{(d+1)}}{L^{d-2}}\hat{\mathcal{S}}_{Im}^T
    =-i\, \frac{2r_0^{\left(1-\frac{1}{2 \beta }\right)z}}{z}\frac{ \sqrt{\pi }  \beta \cot \left(\pi \beta\right) \Gamma \left(-\frac{1}{2}+\beta\right)}{\Gamma \left(\beta\right)}~,
\end{align}
\begin{align}
    \frac{4G_N^{(d+1)}}{L^{d-2}}\hat{S}_{Re}^T = \frac{4G_N^{(d+1)}}{L^{d-2}}\left(\mathcal{S}^T_{Re}-\mathcal{S}^T_{discon}\right)
    =\frac{2 r_0^{\left(1-\frac{1}{2 \beta }\right) z}}{z}\frac{\beta \, \Gamma (1-\beta ) \Gamma \left(\beta -\frac{1}{2}\right)}{\sqrt{\pi }}~.
\end{align}
Utilizing \eq{r0_x1_slice0}, $\hat{\mathcal{S}}_{Im}^T$ and $\hat{\mathcal{S}}_{Re}^T$ can be written in terms of the total subsystem length $T$ as follows
\bea
    &&\frac{4G_N^{(d+1)}}{L^{d-2}}\hat{\mathcal{S}}_{Im}^T
    =i \, f\left( \frac{z}{2 (d-2) z+2}\right)z^{2-d-\frac{1}{z}}\left(\frac{1}{T}\right)^{d-3+\frac{1}{z}}~, \label{Sim_x1slice}\\
    &&\hat{\mathcal{S}}_{Re}^T=i\sec \left( \frac{\pi  z}{2 (d-2) z+2} \right) \hat{\mathcal{S}}_{Im}^T
    \label{Sre_x1slice}
\eea
where $f(\beta)$ is a constant given by
\begin{align}\label{f_alpha_z}
   f(\beta):=-\frac{2 \beta}{2\beta -1}\left(\frac{2 \sqrt{\pi }\, \Gamma (1-\beta)}{\Gamma \left(\frac{1}{2}-\beta \right)}\right)^{\frac{1}{2 \beta}}\left(\sec (\pi  \beta \right)-1)^{\frac{1}{2 \beta }-1}~.
\end{align}
Looking at \eq{Sre_x1slice}, we note that
the two expressions $\hat{\mathcal{S}}_{Re}^T$ and $\hat{\mathcal{S}}_{Im}^T$ differ by a factor that depends on $z$ and $d$. Moreover, as a consistency check, we note that in the isotropic limit $z\rightarrow 1$ the tEE for the different constant space slices along $x$ given by \eq{Sim_x1slice} and \eq{Sre_x1slice} and along $y$ given by \eq{Sim_yslice} and \eq{Sre_yslice}, become equal. Furthermore, in the large dimension limit,
 $|\hat{\mathcal{S}}^T_{Im}|\approx  |\hat{\mathcal{S}}^T_{Re}|$ establishing therefore a property that is valid even in the anisotropic theories. The anisotropic effects in the large $d$-limit, become subleading since in our anisotropic theory we have fixed $d_2=1$.

\subsection{Temporal Entanglement Entropy localized on \texorpdfstring{$y$} ~ direction}\label{aniso_temporal}

In this subsection, we study the temporal entanglement entropy for the Lifshitz-like anisotropic theory. We first consider a constant $y$ slice to analyze the behavior of the extremal surface and subsequently obtain the temporal entanglement entropy. The analysis in this section is equivalent to the AdS with a radius $R^2=1/z^2$ as analyzed in section \ref{conformal}. Nevertheless, we briefly present it.

The equation of motion for the extremal surface is given by
\begin{align}\label{EOM_temp_yslice}
    \tau^{\prime}(r)= \frac{r^{d\, z-1} r_0^{-(d-1)z }}{\sqrt{1-\left(\frac{r}{r_0}\right){}^{2 (d-1) z}}}~,
\end{align}
where we note that the extremal surface can extend to the boundary, i.e. $r<r_0$, when
\begin{align}\label{condition1_temp_yslice}
    (d-1) z>0~,
\end{align}
which effectively constrains $z>0$. Furthermore, notice that if  $z>1/d$, the $\tau^{\prime}(r)$ vanishes at the boundary of the theory and for convenience we work in this parametric regime.

The subsystem length in the Euclidean time direction $\tau$ can be obtained by integrating the expression given in \eq{EOM_temp_yslice} with respect to the bulk direction $r$ and subsequently multiplied by two as
\begin{align}
    T= \frac{2 \sqrt{\pi }\, \Gamma \left(\alpha +\frac{1}{2}\right) }{z\, \Gamma (\alpha )}\,r_0^z~.
\end{align}
Inverting the above relation, the turning point $r_0$ of the extremal surface can be expressed in terms of the subsystem length $T$ as
\begin{align}\label{r0_temp_yslice}
    r_0=\left(\frac{z\, \Gamma (\alpha )}{2 \sqrt{\pi }\, \Gamma \left(\alpha +\frac{1}{2}\right)}\,T\right)^{1/z}~.
\end{align}
The area of the surface extending from the turning point $r_0$ to the boundary is given by \eq{genSint_tau} and reads
\begin{align}
    \frac{4G_N^{(d+1)}}{L^{d-2}}\mathcal{S}^{\tau}
    =2\int_{0}^{r_0} dr~ \frac{r^{-(d-2)z-1}}{\sqrt{1-\left(\frac{r}{r_0}\right){}^{2 (d-1) z}}}~.
\end{align}
As usual, there are UV divergences in the temporal entanglement entropy which can be ``renormalized"   by subtracting the area of the disconnected straight surfaces that extend from the boundary  to the deep IR region as
\begin{align}
    \frac{4G_N^{(d+1)}}{L^{d-2}}\hat{\mathcal{S}}^{\tau}=\frac{4G_N^{(d+1)}}{L^{d-2}}\prt{\mathcal{S}^{\tau}-\mathcal{S}_{discon}^{\tau}}
    =\frac{2 \alpha\sqrt{\pi } \,  \Gamma \left(\alpha -\frac{1}{2}\right) }{z\, \Gamma (\alpha )}\,r_0^{\left(1-\frac{1}{2 \alpha }\right) z}~.
\end{align}
Utilizing \eq{r0_temp_yslice}, we can rewrite the above expression in terms of the subsystem length $T$ as follows
\begin{align}\label{S_temp_yslice}
    \frac{4G_N^{(d+1)} z^{d-1}}{L^{d-2}}\hat{\mathcal{S}}^{\tau}&=g\left( \frac{1}{2(d-1)} \right)\left(\frac{1}{T}\right)^{\left(d-2 \right)}
       = \frac{2^{d-1} \pi ^{\frac{d-1}{2}} }{2-d}\left(\frac{\Gamma \left(\frac{d}{2 (d-1)}\right)}{\Gamma \left(\frac{1}{2 (d-1)}\right)}\right)^{d-1}\left(\frac{1}{T}\right)^{\left(d-2 \right)}~.
\end{align}
where $g(\alpha)$ with $\alpha:=\prt{2 (d-1)}^{-1}$ given by
\begin{align}\label{g_alpha_z}
    g(\alpha)=\frac{2 \alpha }{2 \alpha -1}\left(\frac{2 \sqrt{\pi }\, \Gamma \left(\alpha +\frac{1}{2}\right)}{ \Gamma (\alpha )}\right)^{\frac{1}{2 \alpha }}~.
\end{align}
As expected, the scaling power on $T$ does match with the tEE \eq{Sre_yslice}. Notice that the $z$-dependence appears on the left hand side of \eq{S_temp_yslice}, this is the tEE in the AdS with a radius $R^2=1/z^2$. The temporal entanglement entropy in this case agrees with the tEE before Wick rotation computed for a strip subsystem in the Poincar\'e $EAdS_{d+1}$ geometry of \cite{Doi:2023zaf}.

\subsection{Temporal Entanglement Entropy localized on \texorpdfstring{$x_1$} ~ direction}\label{temporal_aniso_x1}
Let us now work with the temporal entanglement entropy with a constant $x_1$ slice and study the anisotropic exponent that affects this non-local observable.

The equation of motion for the extremal surface is given by
\begin{align}\label{EOM_temp_x1slice}
    \tau^{\prime}(r)= \frac{r^{(d-1) z} r_0{}^{-(d-2) z-1}}{\sqrt{1-\left(\frac{r}{r_0}\right)^{2 (d-2) z+2}}}~,
\end{align}
where the extremal surface exists for $r<r_0$ only when $(d-2) z>-1.$
Positive values of $z$ imply that $\tau^{\prime}(r)$  vanishes at the boundary. For convenience, we work for $A_2: z\ge1$ as suggested by \eq{case2A} and NEC. The subsystem length in the Euclidean time direction $\tau$ is given by the integration of \eq{EOM_temp_x1slice}
\begin{align}
    T= \frac{2 \sqrt{\pi } \,\Gamma \left(\beta +\frac{1}{2}\right)}{z\, \Gamma (\beta )}\, r_0^z~,
\end{align}
where we remind that $\beta:=\frac{z}{2 (d-2) z+2}$.
Subsequently, the area of the extremal surface is computed by utilizing \eq{genSint_tau}
\begin{align}
    \frac{4G_N^{(d+1)}}{L^{d-2}}\mathcal{S}^{\tau}
    =2\int_{0}^{r_0} dr~ \frac{r^{-(d-3)z-2}}{\sqrt{1-\left(\frac{r}{\text{r0}}\right)^{2 (d-2) z+2}}}~.
\end{align}
As earlier, we can ``renormalize" the divergences by subtracting from the area the two infinitely disconnected surfaces as
\begin{align}
    \frac{4G_N^{(d+1)}}{L^{d-2}}\hat{\mathcal{S}}^{\tau}
    =\frac{2 \beta\sqrt{\pi }\, \Gamma \left(\beta -\frac{1}{2}\right) }{z\, \Gamma (\beta )}\,r_0^{\left(1-\frac{1}{2 \beta }\right) z}~,
\end{align}
and obtain the finite expression in terms of the subsystem length $T$ as follows
\begin{align}\label{S_temp_xslice}
	\frac{4G_N^{(d+1)}}{L^{d-2}}\hat{\mathcal{S}}^{\tau}&
		=g\left(\frac{z}{2 (d-2) z+2} \right) z^{2-d-\frac{1}{z}}\left(\frac{1}{T}\right)^{d-3+\frac{1}{z}},
\end{align}
where $g(\beta)$ is given by \eq{g_alpha_z}. % with $\alpha$ replaced by $\beta$.

Notice that the temporal entanglement entropy does depend on the direction we choose to localize the subsystem, and the scaling with the subsystem's length $T$ in the present case is different compared to \eq{S_temp_yslice}. There is a non-trivial $z$-dependence, while in the isotropic limit $z=1$,  the two different $x$ and $y$ slicings, produce the same temporal entanglement entropy as expected. In the limit of large dimensions, in the parametric regime imposed by the NEC: $z\ge1$, the temporal entanglement entropy tends to zero.
Also, we note that our results in the isotropic limit agree with the temporal entanglement entropy computed for a strip subsystem in the Poincar\'e $EAdS_{d+1}$ geometry of \cite{Doi:2023zaf}.

\section{Theories with Hyperscaling Violation}
\label{sec_hyp}
In this section, we discuss the tEE and temporal entanglement entropy for theories with hyperscaling violations utilizing holographic techniques and surface properties.
The gravity side of such theories can be characterized by \cite{Dong:2012se}
\begin{equation}\label{hyp_met}
    ds^2_{d+1}=r^{-\frac{2(d-\theta-1)}{d-1}} \big[ -r^{-2(z-1)}dt^2+dr^2+dx_i^2 \big],
\end{equation}
where $i=1,2,\cdot\cdot\cdot,d-1$. The above metric indicates the spatial homogeneity and scale covariance as,
\begin{equation}\label{hyp_met_sym}
x_i\to \lambda\, x_i,~~t\to \lambda^z t,~~r\to \lambda\, r,~~ds\to \lambda^{\th/d}ds,
\end{equation}
where $z$ and $\th$ are the dynamical exponent and the exponent of hyperscaling violation respectively.

By applying the inequalities \eq{nec2} in \eq{nec}, we obtain two independent NEC for the hyperscaling theory as
\be
(z-1) (d-\theta +z-1)\geq 0~,\qquad (d-\theta -1) ((d-1) (z-1)-\theta )\geq 0~, \la{hyscanec}
\ee
which confine the parametric space $\prt{z,\,\th}$. The AdS spacetime is included in the acceptable range of parameters for $z=1,~\th=0$, as well as the scale invariant Lifshitz theory for $\th=0$, $z=1$.

An extra potential condition to the parameters comes from the black hole solution with hyperscaling violation, by requiring positive specific heat and thermodynamic stability \cite{Dong:2012se} as
\be \la{stable}
\frac{d-\th-1}{z}>0~.
\ee
We will consider the conditions \eq{hyscanec} and \eq{stable} as the minimal necessary ones needed to constrain the parametric space in order to have a natural theory. In fact, the stability condition \eq{stable} applied on top of the NEC \eq{hyscanec} excludes only an additional small area of the parametric space. Let us work here by considering all of these three conditions as desirable for a natural theory.

\subsection{Timelike Entanglement Entropy}\label{hyp_timelike}

We discuss the properties of the spacelike and timelike surfaces and their dependence on the parameters of the theory in our parametrization.

\subsubsection{Timelike Entanglement Entropy Surface Properties}\label{hyp_cov}

The equations of motion in \eq{gentprime} for the two extremal surfaces read as
\begin{align}
    t^\prime=\frac{r^{z-1}}{\sqrt{1+s\left(\frac{r_0}{r}\right)^{2 (d-2+z-\th)}}}~.
\end{align}
As in the previous sections we note that when  $d-2+z-\th>0$, a timelike surface exists for $r>r_0$ extending from the turning point $r_0$ to the infinite $r$.  The extremal surfaces have a non-vanishing and equal $t^\prime_{Im}=t^\prime_{Re}$ in the large $r$ regime where they are merged.
When $d-2+z-\th<0$ the timelike surface extends from $r_0$ to $r=0$. We discuss below more systematically the types of surfaces and their properties depending on the parametric space of the theory.

The quantities \eq{pptt} and \eq{ii}, for the metric  \eq{hyp_met} read
\begin{align}\label{cov_hyp}
|\cT|^2&=-\frac{r^{2 z-\frac{2 \theta }{d-1}}}{1+s\left(\frac{r}{r_0}\right)^{2 (d-2+z-\th)}}~,\qquad
|\cP|^2=-r^{-2\prt{z+1-\frac{2}{d-1}\th}}|\cT|^2~
    \\
    |I_1|^2&= -s  \left(\frac{r}{r_0}\right)^{2\prt{d-2+z-\th}}~,\qquad
    |I_2|^2=1-|I_1|^2~,
\end{align}
where $|\cT|^2=-r^{2\prt{z+1-\frac{2}{d-1}\th}}|\cP|^2$, is positive   for timelike surfaces.
Since hyperscaling violation theory has two scaling parameters, in order to reduce the complication of presentation, let us consider in advance the desirable property of logarithmic or monotonically decreasing norm of tEE with the interval $T$ . A  dimensional analysis for the tEE  gives
\be \la{Shysca}
\hat{\mathcal{S}}\sim T^{-\frac{d-2-\th}{z}}~,
\ee
therefore we require
\be \la{Shysca1}
\frac{d-2-\th}{z}\ge 0~,
\ee
where the saturation corresponds to the logarithmic behavior of the tEE.
Note that even when we impose the NEC and the thermodynamic stability of the theory, there is space for a non-monotonic behavior of $\hat{\mathcal{S}}$ with respect to $T$.  Therefore this is an extra condition already which splits the parametric plane $(z, \th)$,  with a straight line setting the boundary of the accepted parametric regime.

\begin{figure}[t]
	\begin{minipage}[t]{0.5\textwidth}
		\begin{flushleft}
			\centerline{\includegraphics[width=80mm]{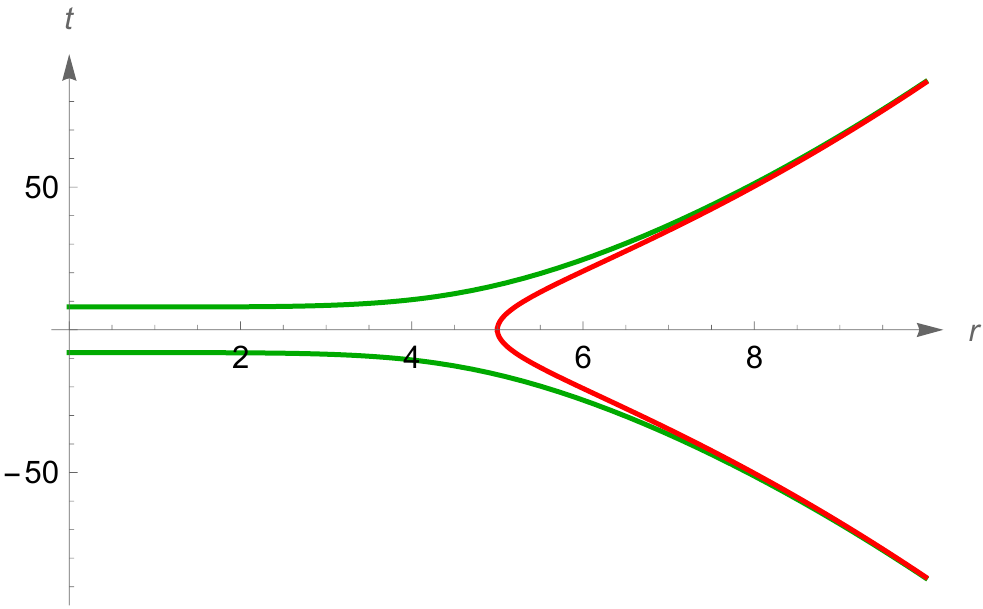}}
			\caption{This surface corresponds to the case $\mathbf{A_0}$, \eq{caseah}.
            The equation of motion for the spacelike hypersurfaces $\Sigma_{Re}$ follow the boundary conditions $t^{\prime}_{Re}|_{r\rightarrow 0}=0$ and $t^{\prime}_{Re}|_{r\rightarrow \infty}=\infty$, whereas for the timelike hypersurface $\Sigma_{Im}$, the boundary conditions are given by $t^{\prime}_{Im}|_{r\rightarrow r_0}=\infty$ and $t^{\prime}_{Im}|_{r\rightarrow \infty}=\infty$. The red curve in the plot describes the timelike surfaces and the green curves are for spacelike ones. We choose $d=4$ and the turning point of the timelike surface is fixed at $r_0=5$. These choices of $d$ and $r_0$ are same for all the neighboring figures of this section. The other parameters here are set to be $z=2$ and $\th=1/2$.}\label{fig:hysca_A0}
		\end{flushleft}
	\end{minipage}
	\hspace{0.3cm}
	\begin{minipage}[t]{0.5\textwidth}
		\begin{flushleft}
			\centerline{\includegraphics[width=80mm ]{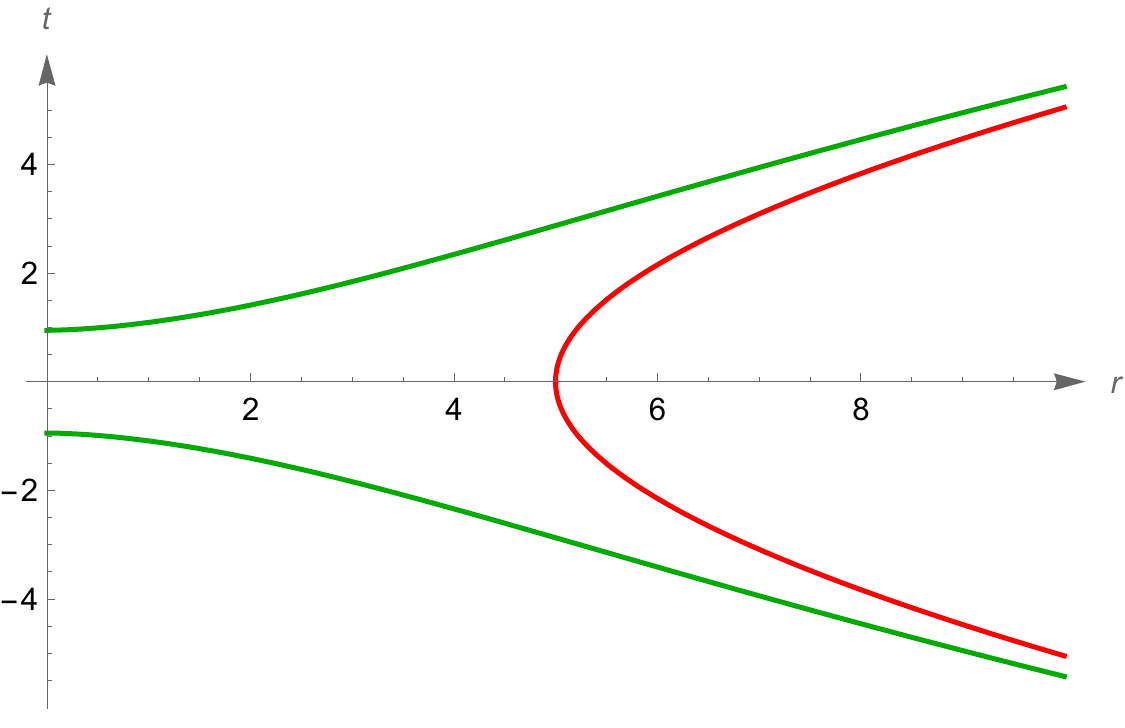}}
			\caption{This surface corresponds to the case $\mathbf{A_1}$, \eq{caseah}.
            For the spacelike surfaces we have
            $t^{\prime}_{Re}|_{r\rightarrow 0}=0=t^{\prime}_{Re}|_{r\rightarrow \infty}$, whereas for the timelike surfaces the boundary conditions are given by $t^{\prime}_{Im}|_{r\rightarrow r_0}=\infty$ and $t^{\prime}_{Im}|_{r\rightarrow \infty}=0$. Here we consider $z=2/5$ and $\th=1.1$.}
			\label{fig:hysca_A1}%\vspace{3.5cm}
		\end{flushleft}
	\end{minipage}
\end{figure}

\begin{figure}[t]
	\begin{minipage}[t]{0.5\textwidth}
		\begin{flushleft}
			\centerline{\includegraphics[width=80mm ]{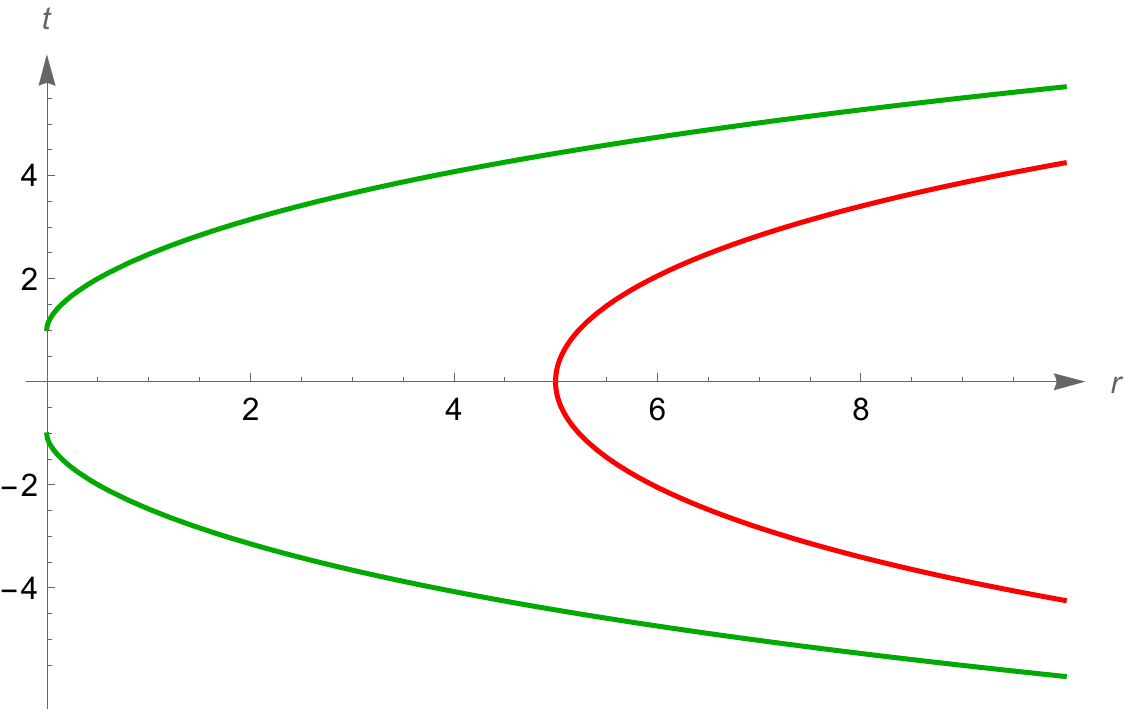}}
			\caption{This surface corresponds to the case $\mathbf{A_2}$ of \eq{caseah}.
            The equation of motion for $\Sigma_{Re}$ follows the boundary conditions $t^{\prime}_{Re}|_{r\rightarrow 0}=\infty$ and $t^{\prime}_{Re}|_{r\rightarrow \infty}=0$, whereas for $\Sigma_{Im}$, the boundary conditions are given by $t^{\prime}_{Im}|_{r\rightarrow r_0}=\infty$ and $t^{\prime}_{Im}|_{r\rightarrow \infty}=0$. Here we have $z=1/10$ and $\th=1.6$.}
			\label{fig:hysca_A2}
		\end{flushleft}
	\end{minipage}
	\hspace{0.3cm}
	\begin{minipage}[t]{0.5\textwidth}
		\begin{flushleft}
			\centerline{\includegraphics[width=80mm]{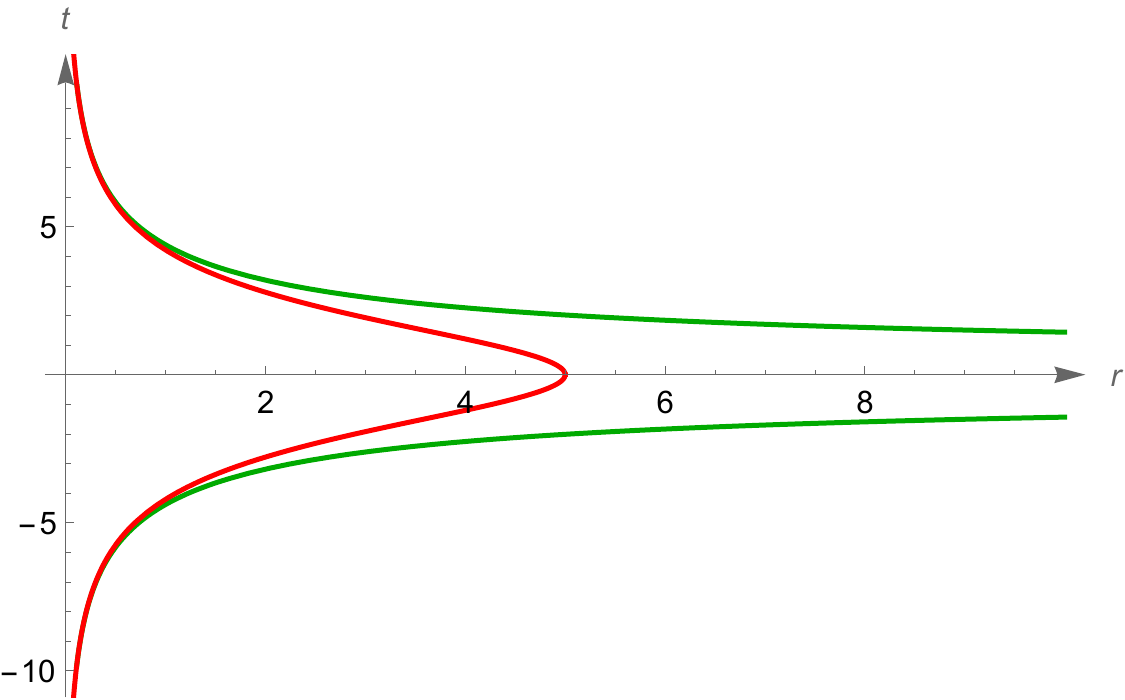}}
			\caption{This surface corresponds to the case $\mathbf{B}$ of \eq{casebh}. The boundary of the geometry is at $r=0$ fixed by $\th\le d-1$. The equation of motion for $\Sigma_{Re}$ follows the boundary conditions $t^{\prime}_{Re}|_{r\rightarrow 0}=\infty$ and $t^{\prime}_{Re}|_{r\rightarrow \infty}=0$. For $\Sigma_{Im}$, the boundary conditions are given by $t^{\prime}_{Im}|_{r\rightarrow r_0}=\infty$ and $t^{\prime}_{Im}|_{r\rightarrow 0}=\infty$. Here we consider $z=-1/5$ and $\th=2.5$.}\label{fig:hysca_B}
		\end{flushleft}
	\end{minipage}
	
\end{figure}

Let us split the following discussion by the position of the boundary of the theory. For
\be \la{boundary1}
\th\le d-1~,
\ee
the boundary of the theory is at $r=0$. Then we classify the cases depending on the scaling of the $I_1$ and $t'$:
\bea\nn
&&\mathbf{A:}~ d-2+z-\th\ge0~\Rightarrow \mathbf{A_0}: \th\le d-2~,~ z\ge1~;\quad\mathbf{A_1}:~\th\le 2z-3+d~,~ z<1, \\&&\mathbf{A_2}:~ 2z-3+d<\th\le d-2~,~ z>0~,  \la{caseah}\\  &&\mathbf{B:}~ d-2+z-\th<0~\Rightarrow ~d-2\le\th\le d-1~,~ z<0,\la{casebh}
\eea
where we have used \eq{Shysca1} and \eq{boundary1} already. Indicative surfaces are plotted as follows $A_0$: Figure \ref{fig:hysca_A0}, $A_1$: Figure \ref{fig:hysca_A1}, $A_2$: Figure \ref{fig:hysca_A2}, $B$: Figure \ref{fig:hysca_B}, where in captions we discuss their basic properties.

The surfaces $A_0$ (Figure \ref{fig:hysca_A0}), have the same behavior as in the conformal theory. The timelike surface extends from the turning point in the bulk to the deep IR. The derivative $t'$ of the spacelike surfaces vanishes at the boundary of the theory and is monotonically increasing from the UV to the IR region. For both timelike and spacelike surfaces the dervative $t^{\prime}$ diverges at the deep IR. Besides, for the timelike surface, $t^\prime$ diverges at the turning point $r=r_0$.
These are what we call as conventional surfaces, the surface that has common behavior with the surfaces corresponding to conformal field theories.

\begin{figure}[!t]
	\centerline{\includegraphics[width=80mm]{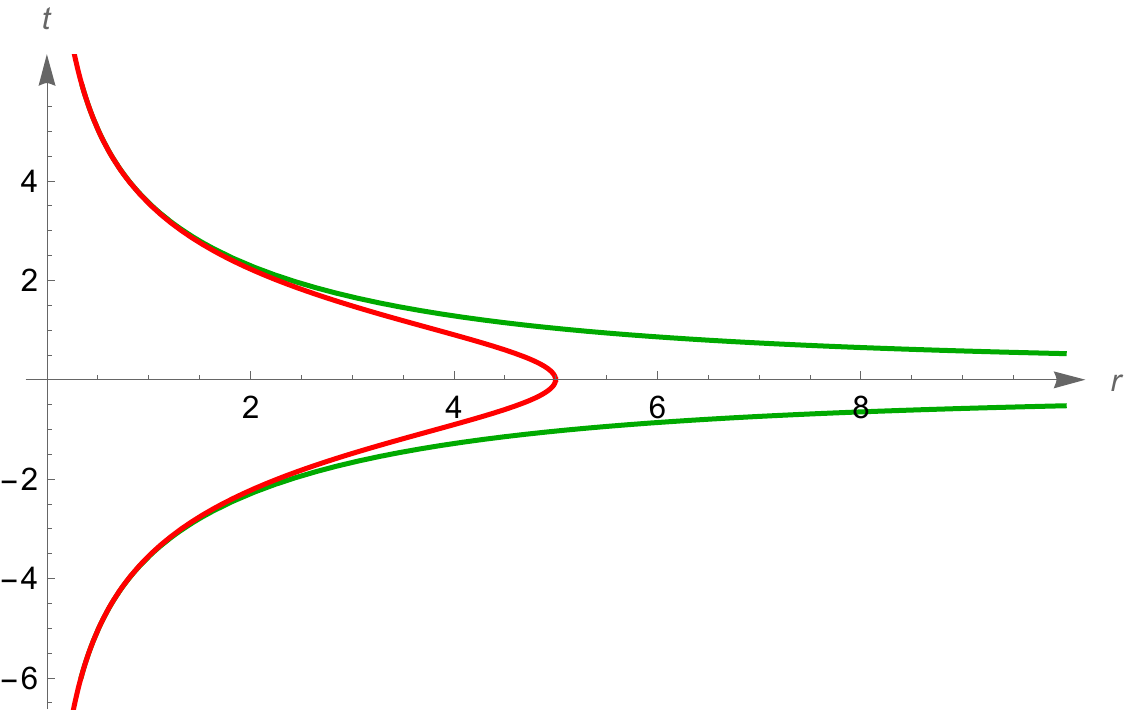}}
	\caption{This surface corresponds to the case $\mathbf{C}$ of \eq{casech}. The boundary of the geometry here is at $r=\infty$ fixed by $\th> d-1$. The equation of motion for $\Sigma_{Re}$ follows the boundary conditions $t^{\prime}_{Re}|_{r\rightarrow 0}=\infty$ and $t^{\prime}_{Re}|_{r\rightarrow \infty}=0$. For $\Sigma_{Im}$, the boundary conditions are given by $t^{\prime}_{Im}|_{r\rightarrow r_0}=\infty$ and $t^{\prime}_{Im}|_{r\rightarrow 0}=\infty$. Here we consider $z=-1/5$ and $\th=3.1$.}\label{fig:hysca_C}
\end{figure}

\begin{figure}[!t]
	\begin{minipage}[t]{0.5\textwidth}
		\begin{flushleft}
			\centerline{\includegraphics[width=80mm]{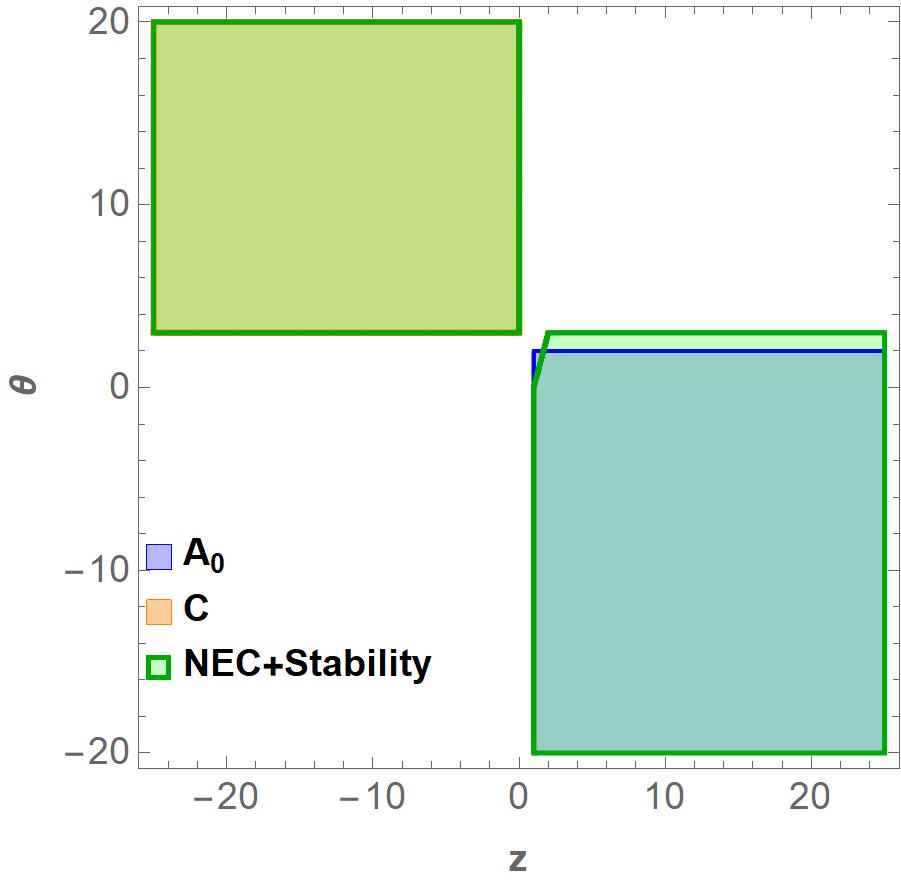}}
			\caption{The NEC \eq{hyscanec} for the hyperscaling violation theory together with the thermodynamic stability conditions \eq{stable} are shaded with light green.  The stability conditions exclude an additional minor regime compared to NEC and result to two disconnected regimes. The blue colored regime includes the conditions $A_0$ from \eq{caseah} which constrain the parametric space to be almost identical to the positive $z$-branch of the NEC. In addition, we have the conventional surfaces of the type $C$ \eq{casech}, plotted with orange that fully match with the negative $z$-branch of the NEC and comprise the left top shaded overlapping area. The unconventional surfaces $A_1$, $A_2$ and $B$ have no overlap with the NEC plus stability conditions.}\label{fig:A0h}
		\end{flushleft}
	\end{minipage}
	\hspace{0.3cm}
	\begin{minipage}[t]{0.5\textwidth}
		\begin{flushleft}
\centerline{\includegraphics[width=80mm ]{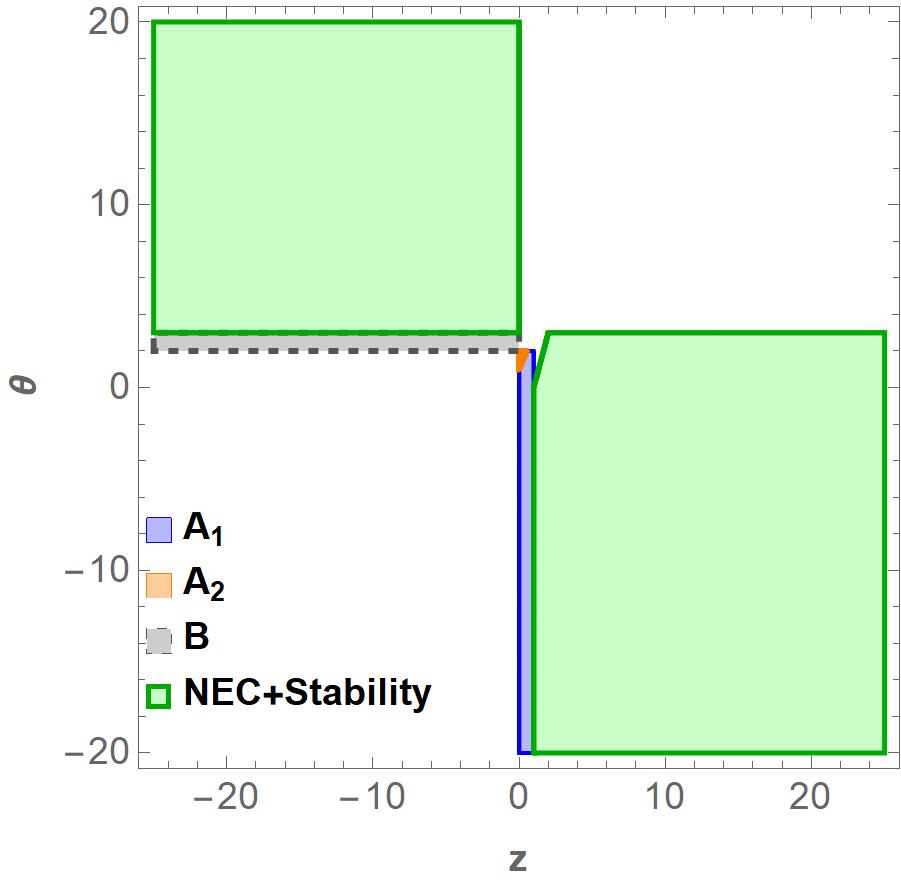}}
			\caption{As in Figure \ref{fig:A0h} the NEC \eq{hyscanec} together with the thermodynamic stability conditions \eq{stable} are shaded with green.  The blue, orange and gray colored regimes correspond to the $A_1$, $A_2$ from \eq{caseah} and $B$ from \eq{casebh} unconventional surfaces respectively. We observe that these surfaces appear for parameters that are outside the accepted parametric regime given by the NEC and the thermodynamic stability. The region plot is made for $d=4$ without loss of generality of the observations mentioned.  }\label{fig:A1h}%\vspace{1.2cm}
		\end{flushleft}
	\end{minipage}
\end{figure}

It is worthy to highlight that the $A_0$ branch, restricts the parametric space to be almost identical to the positive-$z$ branch satisfied by the NEC and the thermodynamic stability (Figure \ref{fig:A0h}).
In fact, for the top small strip-like region on the positive $z$ branch between NEC and $A_0$, where the overlap is imperfect, we expect a better overlap to occur if we allow tEE behaviors that interpolate between logarithmic and linear by relaxing \eq{Shysca1}. However, we expect these phases to be unnatural; therefore, we argue that the correct constraints are as shown in Figure \ref{fig:A0h}.  Moreover, we comment that the tiny triangle region of the positive $z$ that is not fully overlapping in the Figure \ref{fig:A0h} may be excluded by imposing additional conditions, for instance, such that $|\cT|^2\ge|\cT|^2{}_{AdS}$ at the UV, which is essentially the appropriate slope of the NEC there; we have not imposed this extra-fine type of conditions however in this work.

The surfaces in the parametric regime $A_1$ approach the boundary and the deep IR with vanishing $t'$, while the $A_2$ surfaces have infinite $t'$ at the boundary and vanishing at the deep bulk.  Furthermore, we observe that for parametric regime $B$ where $t^\prime$ diverge for both the surfaces at the boundary whereas for the spacelike surface it vanishes in the deep IR region. Interestingly, the unconventional surfaces $A_1$, $A_2$ and $B$ correspond to the unphysical parametric regime $\prt{z,\th}$, as it can be seen in  Figure \ref{fig:A1h}. Notice that the surfaces $A_1$ although they have different behavior in the deep IR compared to the conformal ones,  their boundary conditions are sensible. Nevertheless, the regime of Lifshitz and hyperscaling exponents  that correspond to, is outside the NEC.

In summary for $\th\le d-1$ where the boundary of the theory is at $r=0$, we obtain the conventional form of tEE surfaces for a $\prt{z,\,\th}$-regime that is almost identical to the positive-$z$ branch of the thermodynamic stability together with NEC.

To complete the classification we examine hyperscaling theories with boundary at $r\rightarrow\infty$, which happens for
\be\la{boundary2}
\th> d-1~.
\ee In this case we have only a set of surfaces that satisfy
\be\la{casech}
\mathbf{C:}~ z<0~,
\ee
on top of \eq{Shysca1} and \eq{boundary2}. The timelike surface (Figure \ref{fig:hysca_C}), extends from the turning point $r_0$ to the deep IR for $r<r_0$. Both spacelike and timelike surfaces have diverging
$t'$  at the deep IR and the spacelike has vanishing $t'$ at the boundary of the theory, resembling the behavior of the surfaces of the case $A_0$ and of the conformal theory. Interestingly enough, the surfaces $C$ correspond identically to the negative-$z$ branch of the thermodynamic stability plus NEC, as it can be seen in the Figure \ref{fig:A1h}.

So far we have done the classification of the properties of the surfaces and we have shown that surfaces with conventional behavior restrict the parametric space of the theory to match the stability and NEC. Notice that one could impose further conditions working with \eq{cov_hyp}.  For example, on how fast ${I_{1}}$  vanishes with respect to the conformal surfaces, or investigate the rate of change of $\cP$ and $\cT$ of the hypersurfaces in non-relativistic conformal theory compared to the ones in the conformal theory. Then it is possible that the parametric space of $A_0$ can be reduced even further to be fully inside the NEC regime.

In summary, working in the Poincare coordinate system and in the convenient static gauge parametrization, we provide a classification of the behavior of the timelike and spacelike surfaces that comprise the tEE. The behavior of the surfaces depends on the values of the parameters $(z, \theta)$ responsible for breaking the symmetry of the theory. When the NEC conditions are satisfied, we find that the tEE surfaces in non-relativistic theories have common characteristics with the surfaces of conformal field theories. Moreover, by working in the opposite direction we show that when natural monotonicity conditions are imposed on the tEE, such as requiring a monotonically decreasing tEE with respect to the interval $T$, together with the conformal-like behavior of the hypersurfaces, we find  the parametric space $\prt{z,\th}$ of the theory to be almost identical to the parametric physical space required by the NEC and the thermodynamic stability. Our finding is summarized by the Figures \ref{fig:A0h} and \ref{fig:A1h}.

\subsubsection{Lifshitz Theories: Timelike Entanglement Entropy Surface Properties}

Our analysis becomes considerable simpler when one considers the scale invariant Lifshitz theories $\th=0$. In this case, the tEE reads
\be \la{Slif}
\hat{\mathcal{S}}\sim T^{-\frac{d-2}{z}}~,
\ee
and therefore from \eq{Shysca1} we get
\be \la{Slif1}
z\ge 0~.
\ee
Then the surfaces $B$ \eq{casebh}, $C$ \eq{casech} and $A_2$ \eq{caseah} are outside this regime. The $A_1$ type of surfaces have the same unconventional properties described in the previous section and the only one with the desirable properties is the $A_0$ \eq{caseah} for the parametric space that fully matches with the NEC:
\be
z\ge1~.
\ee

\subsubsection{Computation of Timelike Entanglement Entropy}\label{hyp_tee_comp}

In this subsection, we compute tEE for the strip like subsystem $A$ at the boundary. We first determine the length of the strip like subsystem $A$ in terms of the turning point $r_0$ of the surface $\Sigma_{Im}$. The length of the subsystem corresponding to $\Sigma_{Re}$ can be computed utilizing \eq{gensub} for $c^2>0$, or equivalently \eq{gentprime} for $s=1$. It gives
\begin{align}
    T_{Re}=2\int^\infty_0 t^\prime_{Re}~dr
    =\frac{2 \epsilon_1 ^{-z}}{z}-\frac{2r_0^z}{z}
    \frac{ \Gamma \left(1-\gamma\right) \Gamma \left(\frac{1}{2}+\gamma\right)}{\sqrt{\pi }},
\end{align}
where $\epsilon_1$ is a small number utilized as the IR cutoff which cancels in \eq{r000}, and for presentation reasons we define the $\prt{d,z,\th}$ dependent constant $\gamma:=\frac{z}{2 (d-2+z-\th)}$. Similarly for the timelike surface $\Sigma_{Im}$, its length is given by
\begin{align}
    T_{Im}=2\int^\infty_{r_0} t^\prime_{Im}~dr
    =\frac{2 \epsilon_1 ^{-z}}{z}-\frac{2r_0^z}{z}
    \frac{\sqrt{\pi } \Gamma \left(1-\gamma\right)}{\Gamma \left(\frac{1}{2}-\gamma\right)}~.
\end{align}
The size of the subsystem $A$ can be obtained by considering the difference between $T_{Im}$ and $T_{Re}$ \cite{Afrasiar:2024lsi} as
\begin{align}\label{r000}
    T=T_{Im}-T_{Re}\
    =\frac{2r_0^z}{z}
    \frac{\sqrt{\pi } \left(\sec \left(\pi \gamma\right)-1\right) \Gamma \left(1-\gamma\right)}{\Gamma \left(\frac{1}{2}-\gamma\right)}~.
\end{align}
The real part of tEE is obtained from the area of the spacelike surface $\Sigma_{Re}$ following \eq{genSintRe} and is equal to
\begin{align}\label{TEE_re_hyp}
    \frac{4G_N^{(d+1)}}{L^{d-2}}\hat{\mathcal{S}}^T_{Re}
    =\frac{2 r_0^{\left(1-\frac{1}{2 \gamma }\right) z}}{z} \frac{ \gamma\, \Gamma \left(\gamma-\frac{1}{2}\right) \Gamma \left(1-\gamma\right)}{\sqrt{\pi }}~.
\end{align}
In the above expression, we considered the renormalized area by subtracting the infinite straight surfaces to get rid of the divergences from the UV region. The divergences from the IR region can be avoided by utilizing the bound $d-\th-2>0,$
which is obtained for the case $A_0$ given in \eq{caseah}.
Utilizing \eq{r000}, we express the real part of the tEE as,
\begin{align}\label{TEE_re_hyp_T}
    \frac{4G_N^{(d+1)}}{L^{d-2}}\hat{\mathcal{S}}^T_{Re}&=-f(\gamma) \sec\left(\pi  \gamma\right) z^{-\frac{1}{2\gamma}} \left(\frac{1}{T}\right)^{\frac{1}{2 \gamma }-1}\nn\\
    &= -f\left(\frac{z}{2 (d-2+z-\th)} \right) \sec \left(\frac{\pi  z}{2 (d-2+z-\th)}\right) z^{-\frac{d-2+z-\th}{z}} \left(\frac{1}{T} \right)^{\frac{d-2-\th}{z}}.
\end{align}
where the constant $f(\gamma)$ is given by \eq{f_alpha_z}.

Similar to the real part of tEE, the imaginary part can also be obtained using \eq{genSintIm} and is equal to
\begin{align}\label{TEE_im_hyp}
    \frac{4G_N^{(d+1)}}{L^{d-2}}\hat{\mathcal{S}}^T_{Im}%=2\int^\infty_{r_0} S_{Im}^Tdr
    =i \frac{2 r_0^{\left(1-\frac{1}{2 \gamma }\right) z}}{z}\frac{\sqrt{\pi\, } \gamma \, \Gamma (1-\gamma )}{\Gamma \left(\frac{3}{2}-\gamma \right)}~.
\end{align}
The imaginary part of TEE in terms of the length of the subsystem $T$ as,
\begin{align}\label{TEE_im_hyp_T}
    \frac{4G_N^{(d+1)}}{L^{d-2}}\hat{\mathcal{S}}^T_{Im}&=i\,f(\gamma)\, z^{-\frac{1}{2\gamma}} \left(\frac{1}{T}\right)^{\frac{1}{2 \gamma }-1}
    =i\, f\left( \frac{z}{2 (d-2+z-\th)}\right)z^{-\frac{d-2+z-\th}{z}} \left(\frac{1}{T} \right)^{\frac{d-2-\th}{z}}.
\end{align}
The real and the imaginary parts differ only by a factor $i\sec\prt{\pi\g}$, as
\be
\hat{\mathcal{S}}^T_{Re}=i \sec \left(\frac{\pi  z}{2 (d-2+z-\th)}\right) \hat{\mathcal{S}}^T_{Im}~.
\ee

\subsubsection{Fermi Surface and Timelike Entanglement Entropy}\label{hyp_timelike_log}

We define the systems with Fermi surfaces by the criterion that their entanglement entropy shows a logarithmic violation of the area law. Here we show that Fermi surfaces can be defined via the logarithmic behavior of the real part of the tEE, or the constant $z$-dependent imaginary part.

To do that we note that the analysis for the tEE has to be done as a separate case for $\th=d-2$ i.e. $\gamma=\frac{1}{2}$, since for the integrations of real \eq{TEE_re_hyp_T} and imaginary  \eq{TEE_im_hyp_T} parts of tEE we assumed that $\th$ does not take this particular value.

Utilizing this special value of $\th$, the equations of motion for $\Tilde{\Sigma}_{Im}$ and $\Tilde{\Sigma}_{Re}$ are
\begin{align}
    \tilde{t}^\prime_{Im}=\frac{r^{z-1} r_0^{-z}}{\sqrt{r_0^{-2 z}-r^{-2 z}}},\qquad
    \tilde{t}^\prime_{Re}=\frac{r^{z-1} r_0^{-z}}{\sqrt{r_0^{-2 z}+r^{-2 z}}}.
\end{align}
The length of the subsystems corresponding to the timelike and spacelike surfaces are computed as,
\begin{align}
\tilde{T}_{Re}=\frac{2 \sqrt{\tilde{r}_0^{2 z}+\epsilon_1 ^{-2 z}}}{z}-\frac{2 \tilde{r}_0^z}{z}~,\qquad
\tilde{T}_{Im}=\frac{2 \sqrt{\epsilon_1 ^{-2 z}-\tilde{r}_0^{2 z}}}{z}~.
\end{align}
where we consider $z\ge2-1/(d-1)$ such that we satisfy the NEC \eq{hyscanec} and we have the type $A_0$ surfaces of \eq{caseah} and $\frac{1}{\epsilon_1}\gg r_0$ as $\epsilon_1$ is the IR cutoff that does not appear in the system's total length $\tilde{T}$. Utilizing the above two equations, the length of the subsystem, $T$ can be expressed as,
 \begin{align}
\tilde{T}&=\tilde{T}_{Im}-\tilde{T}_{Re}=\frac{2 \tilde{r}_0^z}{z}~.
\end{align}
The real and the imaginary parts of tEE can be obtained following the same procedure presented in \eq{hyp_tee_comp} as,
\begin{align}\label{TEE_hyp_log}
    \tilde{\mathcal{S}}^T_{Re}
    =\frac{2}{z} \log \left(\frac{z \tilde{T}}{\tilde{\epsilon }}\right),\qquad
    \tilde{\mathcal{S}}^T_{Im}
    =\frac{i \pi }{z}~,
\end{align}
where $\Tilde{\e}=\e^z$ is the redefined UV cutoff. The real part of the tEE shows a logarithmic violation of the tEE area law, signalling the appearance of a Fermi surface in the dual theory. Furthermore, the imaginary component exhibits a consistent, $T$-invariant behavior, indicating the presence of Fermi surfaces in a novel manner. The logarithmic violation term does depend of the scaling $z$ providing additional information on the critical exponent of the Fermi surfaces, compared to the holographic entanglement entropy. The same is true as well for the constant imaginary part. Furthermore, we find that the results in \eq{TEE_hyp_log} correspond to the $2+1$ dimensional holographic dual of $1+1$ dimensional Lifshitz field theory \cite{Basak:2023otu}, as expected from the definition of the systems that exhibit Fermi surface.

\subsection{Temporal Entanglement Entropy }\label{hyp_temporal}
In this subsection, we consider the metric described in \eq{hyp_met} in Euclidean signature to analyze the behavior of the extremal surface at a constant space slice and subsequently compute the temporal entanglement entropy.

The equation of motion corresponding to the extremal surface extending from the boundary $r\rightarrow 0$ to the turning point $r_0$ into the bulk, is given by
\begin{align}\label{EOM_hyp_temp}
    \tau^{\prime}(r)= \frac{r^{d-\theta +2 z-3} r_0^{-d+\theta -z+2}}{\sqrt{1-\left(\frac{r}{r_0}\right)^{2 (d-2+z-\th)}}}~.
\end{align}
The subsystem length considered in the Euclidean time direction $\tau$ can be obtained by integrating \eq{EOM_hyp_temp} with respect to $r$ and subsequently multiplied by two as
\begin{align}\la{r0_hyp_temp0}
    T= 2\int_{0}^{r_0} dr~ \tau^{\prime}(r)= \frac{2 r_0^z}{z} \, \frac{\sqrt{\pi }\, \Gamma \left(\gamma +\frac{1}{2}\right)}{\Gamma (\gamma )}~,
\end{align}
for the parametric space defined by $A_0$ given in \eq{caseah}. For the temporal entanglement entropy, one may again consider the different types of surfaces depending on the parameters of the theory similar to those of the timelike entanglement entropy. The temporal entanglement entropy corresponding to the extremal surface described above can be computed from \eq{genSint_tau} as
\begin{align}
    \frac{4G_N^{(d+1)}}{L^{d-2}}\mathcal{S}^{\tau}
    = 2\int_{0}^{r_0} dr~ \frac{r^{-(d-\theta -1)}}{\sqrt{1-\left(\frac{r}{r_0}\right)^{2 (d-2+z-\th)}}}~.
\end{align}
The above integral contains UV divergence at the boundary $r \rightarrow 0$. The finite part from this integral can be obtained by subtracting the area of the disconnected surface which extends from the boundary $r \rightarrow 0$ to deep into the bulk $r \rightarrow \infty$ as
\begin{align}
    \frac{4G_N^{(d+1)}}{L^{d-2}}\hat{\mathcal{S}}^{\tau}
    =\frac{2 r_0^{\left(1-\frac{1}{2 \gamma }\right) z}}{z} \frac{\sqrt{\pi } \,\gamma\,  \Gamma \left(\gamma -\frac{1}{2}\right)}{\Gamma (\gamma )}~.
\end{align}
Utilizing \eq{r0_hyp_temp0}, we can now rewrite the above expression in terms of the subsystem length $T$ as
\begin{align}
    \frac{4G_N^{(d+1)}}{L^{d-2}}\hat{\mathcal{S}}^{\tau}&=g(\gamma)\, z^{-\frac{1}{2\gamma}} \left(\frac{1}{T}\right)^{\frac{1}{2 \gamma }-1}
    =g\left(\frac{z}{2 (d-2+z-\th)} \right) z^{-\frac{d-2+z-\th}{z}} \left(\frac{1}{T} \right)^{\frac{d-2-\th}{z}},
\end{align}
where the constant $g(\gamma)$ is defined by \eq{g_alpha_z}. % with $\alpha$ replaced by $\gamma$.

As discussed in \eq{temporal_aniso_x1}, we compare the above result of temporal entanglement entropy in the isotropic limit $\theta\rightarrow 0$ and $z\rightarrow 1$ with the corresponding tEE before the Wick rotation $t\rightarrow i\,t$ obtained for a strip subsystem in the Poincar\'e $EAdS_{d+1}$ spacetime described in \cite{Doi:2023zaf}. We found an exact match between the two, which provides another consistency check of our result.

\subsubsection{Fermi Surface and Temporal Entanglement Entropy}\label{hyp_temporal_log}

We show that systems with Fermi surfaces can be defined via the logarithmic behavior of the real part of temporal entanglement entropy. We proceed to compute the temporal entanglement entropy for a specific choice of parameters where $\th=d-2$. The equation of motion for the surface is,
\begin{align}\label{tp_temp_log}
    \tilde{\tau}^\prime&=\frac{r^{2 z-1}}{\sqrt{\tilde{r}_0^{2 z}-r^{2 z}}}.
\end{align}
At the boundary $r=0$, this surface is required to satisfy $\tilde{\tau}^\prime=0$ which happens when the NEC are satisfied.
Integrating \eq{tp_temp_log} we compute the length of the subsystem $A$ as
\begin{align}
\tilde{T}&=\frac{2 \tilde{r}_{0}^z}{z},
\end{align}
where $\tilde{r}_{0}$ is the turning point of the surface.
The temporal entanglement entropy is obtained by computing the area of the surface $\Sigma_{\tau}$,
\begin{align}\label{TEE_re_hyp_temp_log}
    \frac{4G_N^{(d+1)}}{L^{d-2}}\tilde{\mathcal{S}}^\tau%&=2\int^{\tilde{r}_{0}}_0  \tilde{S}^\tau~dr
    =\frac{2}{z} \log \left(\frac{2 \tilde{r}_0^z}{\epsilon ^z}\right).
\end{align}
which can be expressed with the interval $\tilde{T}$ by
\begin{align}\label{kharzeev_tee}
    \frac{4G_N^{(d+1)}}{L^{d-2}}\tilde{\mathcal{S}}^\tau&=\frac{2}{z} \log \left(\frac{z \tilde{T}}{\tilde{\epsilon }}\right),
\end{align}
where $\tilde{\epsilon}=\epsilon^z$ is the redefined UV cutoff.  The temporal entanglement entropy shows a logarithmic violation of the temporal area law, signalling the appearance of a Fermi surface in the dual theory. It does depend of the scaling $z$ providing an additional information of the critical exponent of the Fermi surfaces, compared to the information obtained by entanglement entropy. The temporal entanglement entropy is equal with the real part of the tEE \eq{TEE_hyp_log}. As expected for the systems with Fermi surfaces their temporal entanglement entropy is of the dual $1+1$ dimensional Lifshitz field theory which has been computed also in \cite{Grieninger:2023knz}.

\section{Discussions}

In this article, we explore the holographic tEE in various non-relativistic theories that describe, for example, fixed points with Lifshitz symmetry, hyperscaling violation symmetry, and spatially anisotropic Lifshitz-like symmetries. This is a natural development, given the sensitivity of the tEE to non-relativistic symmetries. We consider the union of the spacelike and timelike extremal surfaces homologous to the timelike region. The real and imaginary components of the tEE correspond to the spacelike and timelike parts of these extremal surfaces, respectively. We also compute the gradient normal vector field, whose norm provides conditions for the spacelike and timelike surfaces in the bulk and where these surfaces meet with each other, illustrating their arrangement for each theory. After determining these surfaces, we compute their extremal areas to derive the real and imaginary components of the tEE, which receive contributions from the spacelike and timelike surfaces respectively. Our approach explicitly agrees with the holographic computations of  \cite{Doi:2023zaf,Afrasiar:2024lsi} when applied to the theories considered in these works,  where the notion of the tEE was previously introduced.

We have summarized most of our results in the introduction already. Here, let us focus more telegraphically on some additional implications. The norm of the unnormalized gradient normal vector field of the tEE hypersurfaces, described in \eq{pptt}, provides conditions for the behavior of spacelike and timelike surfaces in the bulk. For example, in any holographic theory, for timelike and spacelike surfaces to asymptotically end at the same regime, either the norm must be zero at this point, or the sub-volume formed by the time and spatial directions, excluding the direction in which the strip is localized, must vanish.

Moreover, in the Poincare coordinate system and in the static gauge parametrization, we provide a classification of the behavior of the timelike and spacelike surfaces that comprise the tEE, which depends on the values of the parameters $(z, \theta)$ responsible for breaking the symmetry of the system. When the NEC conditions are satisfied, the tEE surfaces in non-relativistic theories share common characteristics with the surfaces of conformal field theories. Additionally, thinking in the opposite direction we show that when extra natural conditions are imposed on the tEE, such as requiring a monotonically decreasing tEE with respect to the interval $T$, together with the conformal-like behavior of the hypersurfaces, the parametric space of the theory is constrained to be almost identical to the one imposed by the NEC and thermodynamic stability conditions of the theory. This implies that the tEE encodes information about a theory’s stability and naturalness. In contrast, entanglement entropy does not contain such information due to its nature involving spacelike intervals.

Our findings strongly suggest that the tEE should be an ideal probe for quantum phase transitions, especially when there is a breaking or emergence of Lorentz symmetries. It is known that quantum critical points with Lifshitz symmetry arise in models of strongly correlated electrons, quantum dimer models on bipartite lattices which exhibit Rokhsar-Kivelson critical points, and in general, Lifshitz symmetry describes diverse quantum critical phenomena \cite{Henley_1997, Ardonne:2003wa, Fradkin_2004, Vishwanath_2004, Ghaemi_2005, PhysRevLett.93.066401, PhysRevLett.61.2376}. Moreover, Lorentz symmetry is often broken in the low-energy phases of matter in solids. Nevertheless, the energy bands of certain materials can exhibit an emergent Lorentz symmetry at low energies in quasiparticle excitations above the ground state. The explicit or spontaneous breaking of this symmetry leads to various phases in the system. A typical example is the Weyl semimetals \cite{weyl1,weyl2,weyl3}, where certain phases include additionally rotational broken symmetry, which we have studied here. Furthermore, holographic models have been proposed to study quantum phase transitions between a topological Weyl semimetal and a trivial semimetal \cite{Landsteiner:2015pdh}, or related phase transitions \cite{Rodgers:2021azg}. Remarkably, the holographic c-function, defined via the classical entanglement entropy, has been proven to be an accurate probe to detect the location of these types of topological quantum critical points \cite{Baggioli:2020cld}. Our results suggest that the tEE has the potential to serve as a novel order parameter in quantum many-body systems, potentially surpassing the entanglement entropy in sensitivity to non-relativistic parameters that measure the degree of broken symmetries.

In the limit of large dimensions, we show that, regardless of the theory we are working on, the real and imaginary parts of tEE tend to become equal in measure. We believe that this is a universal property of the tEE, as it has been noticed in confining theories as well \cite{Afrasiar:2024lsi}.

Moreover, we show that the tEE also serves as a criterion to identify Fermi surfaces. In particular, Fermi surfaces can be defined via the logarithmic behavior of the real part of the tEE, which is analogous to entanglement entropy criteria. Alternatively, Fermi surfaces can be identified via a constant imaginary part of the tEE, with a certain value that depends on the Lifshitz exponent of the theory. Our work provides additional meaning and understanding for the imaginary part of the tEE.

Finally, in our manuscript, we present an extensive analysis of the temporal entanglement entropy in Euclidean signature, which in general differs from the tEE. The temporal entanglement entropy depends on the theory parameters, is sensitive to Lorentz and hyperscaling violation parameters, and, after a Wick rotation on the boundary interval, may be purely imaginary, real, or complex depending on the dimensionality of the system and the values of the non-relativistic exponents of the theory. It is worth noting that the temporal entanglement entropy also hints at the presence of Fermi surfaces in the system.

\section*{Acknowledgment}
The research work of DG is supported by the National Science and Technology Council (NSTC) of Taiwan with the Young Scholar Columbus Fellowship grant 113-2636-M-110-006. The research work of MA and JKB is supported by the NSTC of Taiwan with grant 113-2636-M-110-006.

\appendix

\section{Null Energy Conditions}\label{nec}

We consider the essential pointwise energy conditions in order to have a non-repulsive gravity in anisotropic spacetimes. We review the computation and the result of the NEC of \cite{Chu:2019uoh}. For the study of the energy conditions and without loss of any generality it is more convenient to re-express the generic holographic dual space time of an anisotropic theory metric \eq{genmet} as
\begin{align}\label{aniso_gen}
    ds_{d+1}^2= - e^{2B(\tilde{r})}dt^2 + d\tilde{r}^2 + e^{2 A_1(\tilde{r})}dx_i^2 + e^{2 A_2(\tilde{r})}dy_i^2~,
\end{align}
where the spatial coordinates $x_i$ and $y_i$ extend respectively along $d_1$ and $d_2$ directions such that $d_1+d_2=d-1$.
The NEC  ensure the non-repulsive nature of gravity for the null geodesics \cite{Wald:106274}. Also ensure that the theories do not contain instabilities and avoid superluminal modes in the scalar correlators of the theory \cite{Hoyos:2010at}. The corresponding NEC imposed on the matter fields can be expressed as $T_{\mu\nu}\xi^{\mu}\xi^{\nu} \geq 0~,$
for any null vector $\xi^{\mu}$  satisfying $\xi^{\mu} \xi_{\mu}=0$.
The NEC in terms of the Ricci tensor for the anisotropic theories are written as \cite{Chu:2019uoh}:
\begin{align}
    R^i_i-R^0_0 \geq 0~, \quad R^j_j-R^0_0 \geq 0~, \quad R^{\tilde{r}}_{\tilde{r}}-R^0_0 \geq 0~,
\end{align}
with no summation over the repetitive indices and the $x$ and $y$ coordinates are denoted by the indices $i$ and $j$ respectively.

Applying the above NEC for the anisotropic spacetime described in \eq{aniso_gen} we have three independent conditions as \cite{Chu:2019uoh}:
\begin{align}\la{nec1}
    &\prt{\prt{B'(\tilde{r})-A_1'(\tilde{r})} e^{B(\tilde{r})+k(\tilde{r})}}'\ge 0~,\\\la{nec2}
    &\prt{\prt{B'(\tilde{r})-A_2'(\tilde{r})} e^{B(\tilde{r})+k(\tilde{r})}}'\ge0~,\\%\la{nec3}
    &-d_1 A_1'(\tilde{r})^2-d_2 A_2'(\tilde{r})^2+ B'(\tilde{r}) k'(\tilde{r})-k''(\tilde{r})\ge 0~,
\end{align}
were  $k(r) := d_1 A_1(r) + d_2 A_2(r)$, where the first two equations are written in form of monotonically increasing functions.

An analogous condition to NEC for the null geodesics, appears in the Raychaudhuri equation for the timelike geodesics congruences. In order to secure everywhere an attractive effect on the timelike congruences we need to impose
\be
R_{\m\n} t^\m t^\n\ge0
\ee
for the timelike vectors $t^\m$. These conditions usually refer as strong energy conditions. Here, we focus mainly the implications of the NEC. We have examined the contraction of the tEE hypersurface timelike vectors with the Ricci tensor to obtain explicitly one of the NEC. A more systematic study of the strong energy conditions in relation to the timelike vectors of the tEE hypersurfaces would be interesting.

\bibliographystyle{JHEP}

%\bibliography{timelike}

\providecommand{\href}[2]{#2}\begingroup\raggedright

\end{document}